\documentclass[twocolumn, amsmath, amssymb, aps, prb, floatfix, superscriptaddress
]{revtex4-2}
\usepackage{soul}
\usepackage[T1]{fontenc}
\usepackage{graphicx} 
\usepackage{bm}
\usepackage{hyperref}
\usepackage{physics}
\usepackage{color}
\usepackage{multirow}
\usepackage{orcidlink}
\usepackage{dsfont}
\usepackage{placeins}

\renewcommand{\vb}[1]{\boldsymbol{\mathbf{#1}}}
\newcommand{\Cdag}[1]{c^\dagger_{#1}}
\newcommand{\C}[1]{c^{\phantom{\dagger}}_{#1}}
\newcommand{\up}{\uparrow}
\newcommand{\dn}{\downarrow}

\newcommand{\F}{\mathcal{F}}

\allowdisplaybreaks

\begin{document}
\title{The influence of quantum geometry on the phase boundary and collective excitations of electron liquids and crystals}

\author{Paul Froese \orcidlink{0000-0003-3796-8544}}
\affiliation{
Department of Physics, University of Toronto, Toronto, Ontario M5S 1A7, Canada
}

\author{Mark R. Hirsbrunner \orcidlink{0000-0001-8115-6098}}
\affiliation{
Department of Physics, University of Toronto, Toronto, Ontario M5S 1A7, Canada
}

\author{Yong Baek Kim}
\affiliation{
 Department of Physics, University of Toronto, Toronto, Ontario M5S 1A7, Canada
}

\date{\today}

\begin{abstract}
Recent experiments on multilayer graphene systems have reinvigorated the study of electron crystallization, now with the inclusion of quantum geometry. In this work, we apply time-dependent Hartree-Fock (TDHF) to the $\lambda$-jellium model to analyze the impact that quantum geometry has on the electronic liquid--crystal phase diagram and how it modifies the collective modes and responses of the liquid and crystal phases. In agreement with recent results utilizing neural quantum states, we find that quantum geometry favours electron crystallization, shifting the transition to higher densities. We also study the instabilities revealed by TDHF in the Fermi liquid ground state at low densities, providing insight into the fluctuations driving the crystallization transition. We further find that quantum geometry reduces the dispersion of the plasmon mode and suppresses Friedel oscillations deep in the liquid phase. Resolving the density response in terms of individual orbitals, we find that this suppression is caused by spectral weight transfer to an out-of-phase inter-orbital mode. Finally, we show that an analogous mode that emerges in the crystal phase corresponds to the breathing mode of an emergent real-space pseudospin skyrmion lattice.
\end{abstract}

\maketitle

\section{Introduction}
Rhombohedral $N$-layer graphene (RNG) is a material platform that hosts strong interactions coexisting with non-trivial quantum geometry, supporting the realization of strongly-correlated topological phases of matter. When RNG is nearly aligned with a hexagonal boron nitride (hBN) substrate, an array of integer and fractional quantum anomalous Hall (QAH) phases emerge~\cite{luFractionalQuantumAnomalous2024,suMoiredrivenTopologicalElectronic2025,luExtendedQuantumAnomalous2025,WatersChernInsularors2025,li2025stackingorientationtwistanglecontrolinteger,xieTunableFractionalChern2025,aronsonDisplacementFieldControlledFractional2025,uzan2025hbnalignmentorientationcontrols,huoDoesMoireMatter,zhangMoireEnhancedFlat2025,liuOddChernNumberQuantumAnomalous2026}, in addition to other exotic phenomena that do not require a moir\'e twist, such as signatures of chiral superconductivity~\cite{choiSuperconductivityQuantizedAnomalous2025,Yoon_2026,Seo_2026,dutta2026reconfigurablechiralsuperconductivity,yang2026magneticfieldenhancedgraphenesuperconductivity}. A leading theoretical explanation for the origin of these QAH phases is the formation of an anomalous Hall crystal (AHC), a topological counterpart to a traditional Wigner crystal (WC) that possesses a non-zero Chern number~\cite{zhouFractionalQuantumAnomalous2024,Dong_2024,dongStabilityAnomalousHall2024,Zeng_2025,soejimaTopologicalConstraintCrystalline2025,dongPhononsElectronCrystals2025,desrochersElasticResponseInstabilities2026,soejimaAnomalousHallCrystals2024,tanParentBerryCurvature2024,soejimaJelliumModelAnomalous2025,tanIdealLimitRhombohedral2025,Desrochers_2026}. Recent experiments have also detected signatures of both conventional and metallic Wigner crystals in RNG~\cite{han2026evidencemetallicwignercrystal,zhou2026competingordersdrivenwigner,dong2026crystalscaughtdopingmetallic}. These observations motivate the continued study of electron crystallization in the presence of quantum geometry.

At the mean-field level, the picture of electron crystallization is well-established. Self-consistent Hartree-Fock (HF) calculations have been applied to the microscopic Hamiltonian for RNG twisted on hBN, for which the non-interacting band structure is metallic, but it was found that strong interactions drive the formation of an isolated topological band~\cite{Dong_2024,dongStabilityAnomalousHall2024,zhouFractionalQuantumAnomalous2024}. Motivated by this discovery, a number of simplified models were developed to study electron crystallization in the presence of quantum geometry~\cite{soejimaAnomalousHallCrystals2024,tanParentBerryCurvature2024,soejimaJelliumModelAnomalous2025,Desrochers_2026,tanIdealLimitRhombohedral2025,bernevig2025berrytrashcanmodelinteracting,Cr_pel_2025}. These include the ideal parent band model~\cite{tanParentBerryCurvature2024}, in which a quadratic band of electrons is endowed with constant, uniform Berry curvature, and the $\lambda$-jellium model~\cite{soejimaJelliumModelAnomalous2025}, a minimal extension of conventional jellium that includes Berry curvature, the profile of which is tuned by a length scale $\lambda$. These toy models differ in their underlying degrees of freedom, and can each be understood as limits of a generalized $\lambda_N$-jellium model with $N$ internal degrees of freedom~\cite{Desrochers_2026}. At the mean-field level, the ideal parent band model hosts electron crystal phases at low densities that obtain finite Chern numbers when the Berry curvature is sufficiently strong. The $\lambda$-$N$ jellium models similarly host AHCs, as well as a variety of other liquid and crystal phases enabled by the particular Berry curvature distribution of the model, resulting in a rich mean-field phase diagram.

However, in the conventional two-dimensional electron gas, it is well known that the mean field approximation has severe limitations; the critical density of the liquid-crystal transition predicted by Hartree-Fock is famously off by more than an order of magnitude. Decades of research applying advanced numerical techniques, including quantum Monte Carlo (QMC)~\cite{tanatarGroundStateTwodimensional1989,drummondPhaseDiagramLowDensity2009,azadiQuantumMonteCarlo2024} and sophisticated neural quantum state (NQS) techniques~\cite{smithUnifiedVariationalApproach2024}, have significantly refined our understanding of the transition. However, much is still unknown about the transition even in the absence of Berry curvature, including the nature of any intermediate states~\cite{PhysRevB.67.125205,spivakPhasesIntermediateTwodimensional2004, PhysRevLett.94.056805, RevModPhys.82.1743, Joy_2023,  smithUnifiedVariationalApproach2024,71l6-w8hl, zhou2026competingordersdrivenwigner}. It is therefore crucial to go beyond mean-field to properly understand the liquid--crystal transition in the presence of quantum geometry.

One recent work has done such an investigation, constructing a phase diagram for the $\lambda$-jellium model using state-of-the-art variational NQS wavefunctions~\cite{valentiQuantumGeometryDriven2025}. The NQS calculation finds that the presence of quantum geometry favours electron crystallization, with the critical density of the liquid--crystal transition increasing significantly as $\lambda$ is increased. The proposed explanation for this observation is that the presence of quantum geometry allows electrons to sit nearer to each other in space. For the liquid phase, in which the electrons tend to be closer together than in the crystal phase, this results in a drastic increase in the interaction energy, shifting the boundary in favour of crystallization.

In our work, we investigate the phase diagram from a complementary perspective, using the time-dependent Hartree-Fock (TDHF) method and employing the quasi-boson approximation (QBA). This framework allows us to reach larger system sizes, access the collective modes, and characterize instabilities of mean-field ground states. We apply this method to the $\lambda-$jellium model, first computing the correlation energy to construct the phase diagram. The Fermi liquid (FL) ground state energy we obtain is consistently more negative than that which would be produced by more exact techniques, reflecting the perturbative nature of TDHF, limiting the quantitative accuracy of the technique. Additionally, at the low densities where crystallization occurs, TDHF further suffers because of a mean-field instability in the ground state. Despite this, we obtain a liquid--crystal phase boundary qualitatively similar to that obtained in~\cite{valentiQuantumGeometryDriven2025}, confirming the trend that the presence of quantum geometry increases the critical density. Furthermore, the mean-field instability that causes the breakdown of TDHF provides significant insight into the physics driving the crystallization transition. In the absence of quantum geometry, the two-dimensional electron gas exhibits a well-known instability towards a density wave preceding crystallization~\cite{overhauserSpinDensityWaves1962,overhauserChargedensityWavesIsotropic1978,colonnaCorrelationEnergyExactexchange2014}. Our TDHF results identify an additional instability that emerges only in the presence of quantum geometry, reflecting the tendency of the strong interactions to eject electrons from the Brillouin zone centre, where the quantum geometry is concentrated, out to the Fermi surface. The coincidence of this new instability with the liquid--crystal transition indicates that this type of fluctuation is crucial to competition between the two ground states. 

When applied to the FL at high densities or to the crystal phase at any density for which it is the mean-field ground state, TDHF does not suffer from stability issues and accurately produces the spectrum of collective modes. We utilize the collective modes, which determine the response of the system to external perturbations, to investigate the effect of quantum geometry on the properties of the FL and the various crystal phases hosted by $\lambda-$jellium. In the FL, we find that the quantum geometry significantly suppresses and redistributes the spectral weight of both the density and current responses. This suppression has interesting physical consequences: The dispersion of the collective plasmon mode is reduced, and, in the static limit, Friedel oscillations are completely absent for certain values of $\lambda.$ The origin of this suppression is revealed by computing the orbital-resolved density response, probing the internal structure of the collective excitations. The suppression of the total density response is accompanied by the emergence of an out-of-phase inter-orbital density mode, in which charge oscillates between orbitals while the total density remains fixed. Applying a similar analysis to the crystal phases of the $\lambda-$jellium model, we identify an analogous out-of-phase inter-orbital response that corresponds to the breathing mode of an emergent real-space orbital pseudospin skyrmion lattice~\cite{tanIdealLimitRhombohedral2025,maymann2026skyrmionfractionalcherninsulator}. 

\section{Model and Methods}
\label{sec:MM}
\subsection{$\lambda-$jellium model}
We consider the band-projected $\lambda-$jellium model~\cite{soejimaJelliumModelAnomalous2025}, which takes the form
\begin{equation}
    \begin{gathered}
        \mathcal{H} = \mathcal{H}_0 + \mathcal{H}_\mathrm{int},
        \\
        \mathcal{H}_0 = \sum_{\vb{k}} \varepsilon_{\vb{k}} c^\dagger_{\vb{k}} c^{\phantom{\dagger}}_{\vb{k}}, \quad \mathcal{H}_{\text{int}}= \frac{1}{2A}\sum_{\vb{q} \neq 0}V(\vb{q}) :\hat{\rho}_{\vb{q}} \hat{\rho}_{\vb{-q}}:,
    \end{gathered}
\end{equation}
where $\Cdag{\vb{k}}$ ($\C{\vb{k}}$) creates (annihilates) an electron in the projected band with unbounded momentum $\vb{k},$ and colons indicate normal ordering with respect to the vacuum state. We henceforth work in density-dependent units for which all length scales are given in terms of $r_s a_B = (\pi n)^{-1/2}$, with $r_s$ the dimensionless Wigner-Seitz radius, $a_B = \hbar^2/(me^2)$ the Bohr radius, and $n$ the electron density. In these units, energies are measured in Rydbergs, $1~\mathrm{Ry} = \frac{e^2}{2a_B}$. The kinetic term $\mathcal{H}_0$ describes the single-particle dispersion, given by that of a free electron, $\varepsilon_{\vb{k}} = k^2/r_s^2,$ for which the Fermi momentum is $k_\mathrm{F} = 2$. The area of the system is $L \times L = A$, discretizing momentum space in units of $dk = 2\pi/L$.

The nontrivial quantum geometry of the model is contained in the density operators entering the interaction term, $\mathcal{H}_\mathrm{int}$. Defining the form factors $\mathcal{F}(\vb{k},\vb{k}') = \braket{u_{\vb{k}}}{u_{\vb{k}'}} $, where $\ket{u_{\vb{k}}}$ is the cell-periodic Bloch state, the band projected density operator $\hat{\rho}_{\vb{q}}$ is given by the expression
\begin{align}
    \hat{\rho}_{\vb{q}} = \sum_{\vb{k}} \F(\vb{k} +\vb{q}, \vb{k}) \Cdag{\vb{k} +\vb{q}} \C{\vb{k}}
    .
\end{align}
We take $V(\vb{q})=4\pi/(r_sq)$ to be the unscreened Coulomb interaction, and exclude the $\vb{q} = 0$ term in the interaction Hamiltonian to account for a uniform, neutralizing background charge.

In the $\lambda$-jellium model, the form factor is given by
\begin{align}
    \F(\vb{k},\vb{k}') =  \frac{1 + \lambda^2(k_x + ik_y)(k'_x-ik'_y)}{\sqrt{(1 + \lambda^2k^2)(1+\lambda^2k'^2)}},
\end{align}
where $\lambda$ is a variable length scale. In the $\lambda = 0$ limit, the form factor becomes trivial, recovering the conventional jellium model. For finite $\lambda$, the band acquires a non-uniform Berry curvature distribution, $\Omega(\vb{q}) = 2\lambda^2/(\lambda^2q^2 + 1)^2,$ that integrates to $2\pi$ for any finite value of $\lambda$. The Berry curvature arises from the skyrmionic texture of the occupied band, which is described by the pseudospinor
\begin{equation}
    \ket{u_{\vb{k}}} = \frac{1}{\sqrt{1 + \lambda^2k^2}}\begin{bmatrix}
        1 \\ \lambda(k_x  +ik_y)
    \end{bmatrix}
    .
    \label{eq:spinor}
\end{equation}
We refer to the abstract degrees of freedom represented by the components of the pseudospinor as orbitals, and, for notational simplicity, use the symbols $\up$ and $\dn$ to label them, although they are not physical spins. At $\vb{k}=0$ the pseudospinor is entirely polarized on the first orbital, pointing in the $+\hat{z}$ direction on the Bloch sphere. For finite $\vb{k},$ the pseudospinor cants away from the $\hat{z}$ axis, winding around it as the momentum winds around the origin. At $k = 1/\lambda,$ the pseudospinor points along the equator of the Bloch sphere, and cants downward to the $-\hat{z}$ axis as $k\rightarrow\infty.$ The physics of the model is governed by the interplay of the two parameters, with $r_s$ controlling the density, the size of the Fermi surface, and the strength of interactions, and $\lambda$ controlling the size of the momentum-space skyrmion texture and, consequently, the width of the Berry curvature distribution.

At high densities, $r_s\lesssim2,$ the mean-field ground state of $\lambda-$jellium is a conventional FL, the energy of which is given by
\begin{equation}
    \varepsilon_{\vb{k}}^\mathrm{HF} = \varepsilon_{\vb{k}} - \frac{1}{A} \sum_{\vb{k}'}n_{\vb{k}'} |\mathcal{F}(\vb{k},\vb{k}')|^2 V(\vb{k} - \vb{k}').
    \label{eq:HF_energy}
\end{equation}
We work at zero temperature, such that the occupation is simply given by the step function $n_{\vb{k}} = \theta(k_F - k).$ The magnitude of the form factor is less than or equal to one, $|\mathcal{F}(\vb{k},\vb{k}')|^2 \leq 1,$ and the bound is only satisfied in the conventional jellium model with $\lambda=0.$ As such, the role of quantum  geometry in determining the energetics of the FL is to reduce the amount by which the exchange term lowers the ground state energy. For small $\lambda,$ the skyrmion texture is nearly constant within the Fermi surface, so the impact on the exchange term is minimal. The radius of the skyrmion texture becomes comparable to the Fermi momentum for intermediate values of $\lambda,$ in which case the Fermi sea contains states with pseudospin point along $+\hat{z}$ and $-\hat{z}.$ The form factor connecting states with opposite $\hat{z}$ components of their pseudospin are greatly reduced in magnitude, resulting in a penalty to the exchange energy. In the limit of large $\lambda,$ this penalty becomes extreme for electrons at small momentum. In this case, the system lowers its energy by excluding occupation from a disk centred at $k=0$, producing an annular Fermi sea, with the resulting phase appropriately being dubbed the annular Fermi liquid (AFL)~\cite{soejimaJelliumModelAnomalous2025}.

For $r_s \gtrsim 2$, the Coulomb interaction dominates over the kinetic energy at the mean field level, driving the electrons to spontaneously break translation symmetry and form electronic crystals. For small values of $\lambda$, the ground state is a conventional Wigner crystal, but the non-trivial quantum geometry of the parent band enables two distinct crystalline phases for $\lambda \gtrsim 0.5$. At higher densities, tuning $\lambda \gtrsim 0.5$ drives the WC to an AHC phase wherein the Berry curvature of the parent band is inherited by the valence band of the resulting insulator, endowing it with a Chern number of $C=1$~\cite{soejimaJelliumModelAnomalous2025, dongStabilityAnomalousHall2024}. The WC and AHC both border another phase at lower densities for which occupation of the $\vb{k}=0$ state vanishes, owing to a large energetic penalty similar to that which stabilizes the AFL. In contrast to the AHC, this halo Wigner crystal (HWC) is trivial, having $C=0$~\cite{soejimaJelliumModelAnomalous2025, joyChiralWignerCrystal2025, joy2026wignercrystallizationbernalbilayer}.

\subsection{Time-dependent Hartree-Fock} 
Here we review the TDHF formalism, in which the collective modes of the system are obtained by considering perturbations around the Hartree-Fock ground state~\cite{roweEquationsofMotionMethodExtended1968, ringNuclearManyBodyProblem2004}. The creation operators for the collective modes are approximated as superpositions of elementary particle-hole operators, the energies and eigenfunctions of which are obtained by solving the equations of motion. We employ the quasi-boson approximation to close the equations of motion, asserting that the elementary particle-hole operators obey bosonic commutation relations~\cite{ringNuclearManyBodyProblem2004, Schuck_2021}. This approach provides results comparable to state of the art QMC methods when applied to the unpolarized two-dimensional electron gas~\cite{wolfQuasibosonApproximationYields2024a}, and has been broadly applied in the AHC literature to study the stability of the AHC ground state and its phonon collective modes~\cite{jainElementaryExcitationsMelting2025, huoDoesMoireMatter, kwanMoireFractionalChern2025,  dongPhononsElectronCrystals2025, desrochersElasticResponseInstabilities2026, hirsbrunnerTopologicalPhononsAnomalous2026a,miao2026retardedinteractionoppositechiral, ge2026nematicwignercrystalsrhombohedral}.

In TDHF, one considers the following ansatz for neutral collective modes,
\begin{align}
    Q_{\nu \vb{q}}^\dagger = \sum_{\vb{k} \in K_q} X_{v\vb{q}}(\vb{k})b^\dagger_{\vb{q}}(\vb{k}) 
    -
    \sum_{\vb{k} \in K_{-q}}Y_{v\vb{q}}(\vb{k})b^{\phantom{\dagger}}_{-\vb{q}}
    ,
\end{align}
where we have introduced the elementary particle-hole creation operator $b^\dagger_{\vb{q}}(\vb{k}) = \Cdag{\vb{k}+\vb{q}}\C{\vb{k}}$. The domains over which $\vb{k}$ is summed are defined as $K_{\pm q} = \lbrace \vb{k} | \varepsilon_{\vb{k}} < \varepsilon_{\mathrm{F}} < \varepsilon_{\vb{k} \pm \vb{q}} \rbrace$, such that the first term creates a particle-hole excitation at momentum $+\vb{q}$, while the second term annihilates a particle-hole excitation at momentum $-\vb{q}$. Due to the isotropy of the system, $K_q$ and $K_{-q}$ contain the same number of momentum points, which we denote as $|K_{\pm q}| = M_q$.  The many-body ground state $\ket{0}$ is defined as that which is annihilated by all such operators, namely
\begin{align}
    \label{eq:Q_def}
     Q^{\phantom{\dagger}}_{\nu \vb{q}}  \ket{0} = 0
     .
\end{align}

To proceed, we construct the equations of motion for $Q_{\nu\vb{q}}^\dagger,$
\begin{align}
    [\mathcal{H},Q^\dagger_{\nu\vb{q}}] \ket{0} =  \omega_{\nu \vb{q}}  Q^\dagger_{\nu\vb{q}} \ket{0}
    ,
\end{align}
and recast them using (\ref{eq:Q_def}) as
\begin{align}  
    \mel{0}{[Q^{\phantom{\dagger}}_{\nu \vb{q}},[\mathcal{H},Q_{\nu\vb{q}}^\dagger]]}{0}
    =
     \omega_{\nu \vb{q}} \mel{0}{[Q^{\phantom{\dagger}}_{\nu\vb{q}}, Q_{\nu\vb{q}}^\dagger]}{0}.
\end{align}
We employ the quasi-boson approximation to evaluate these commutators, which amounts to replacing $\ket{0}$ with $\ket{\mathrm{HF}}$. This produces the following generalized eigenvalue equation,
\begin{align}
 \label{eq:TDHF_matrix}
    S_{\vb{q}}
    \begin{bmatrix}
        X_{\nu \vb{q}} \\
        Y_{\nu \vb{q}}
    \end{bmatrix}
    =
    \omega_{\nu \vb{q}}
    \Sigma
    \begin{bmatrix}
        X_{\nu\vb{q}} \\
        Y_{\nu \vb{q}}
    \end{bmatrix}
    ,
\end{align}
with 
\begin{align}
    S_{\vb{q}} =
    \begin{bmatrix}
        \hat{A}_{\vb{q}} & \hat{B}_{\vb{q}} \\
        \hat{B}^*_{-\vb{q}} & \hat{A}^*_{-\vb{q}}
    \end{bmatrix},
    \quad
    \Sigma = \begin{bmatrix}
        \mathds{1} & 0 \\
        0 & - \mathds{1}
    \end{bmatrix}.
\end{align}
Here we have introduced the vectors
\begin{equation}
    X^T_{\nu \vb{q}} = \left[
    X_{\nu\vb{q}}(\vb{k}_1),
    X_{\nu\vb{q}}(\vb{k}_2),
    \dots ,
    X_{\nu\vb{q}}(\vb{k}_{M_q})
    \right]
\end{equation}
 and 
 \begin{equation}
     Y^T_{\nu \vb{q}} = \left[
    Y_{\nu\vb{q}}(\vb{k}'_1),
    Y_{\nu\vb{q}}(\vb{k}'_2),
    \dots,
    Y_{\nu\vb{q}}(\vb{k}'_{M_q})
    \right],
 \end{equation}
 with $\vb{k}_i$ and $\vb{k}'_i$ enumerating the momenta in $K_{q}$ and $K_{-q},$ respectively. The identity operator is written as $\mathds{1}$, and the matrix elements of $\hat{A}_{\vb{q}}$ and $\hat{B}_{\vb{q}}$ are 
\begin{equation}
    \begin{aligned}
        A_{\vb{q}}(\vb{k},\vb{k}')
        &=
        \delta_{\vb{k}\vb{k}'}(\varepsilon^\mathrm{HF}_{\vb{k} + \vb{q}} - \varepsilon_{\vb{k}}^\mathrm{HF})
        \\
        &+
        \frac{1}{A}V(\vb{q})\mathcal{F}(\vb{k}+\vb{q},\vb{k})\mathcal{F}(\vb{k}',\vb{k}' + \vb{q})
        \\
        &-
        \frac{1}{A}V(\vb{k} - \vb{k}')\mathcal{F}(\vb{k}+\vb{q},\vb{k}'+\vb{q})\mathcal{F}(\vb{k}',\vb{k})
    \end{aligned}
\end{equation}
\begin{equation}
    \begin{aligned}
        B_{\vb{q}}(\vb{k},\vb{k}')
        &=
        \frac{1}{A}V(\vb{q})\mathcal{F}(\vb{k}+\vb{q},\vb{k})\mathcal{F}(\vb{k}'-\vb{q},\vb{k}')
        \\
        &-
        \frac{1}{A}V(\vb{k} + \vb{q} - \vb{k}')\mathcal{F}(\vb{k}+\vb{q},\vb{k}')\mathcal{F}(\vb{k}'-\vb{q},\vb{k})
    \end{aligned}
\end{equation}
where $\vb{k},\vb{k}' \in K_q$ for $\hat{A}_{\vb{q}}$, whereas $\vb{k} \in K_q$ and $\vb{k}' \in K_{-q}$ for $\hat{B}_{\vb{q}}$. We solve \eqref{eq:TDHF_matrix} numerically, yielding the collective mode spectrum, $\omega_{\nu\vb{q}},$ and corresponding eigenstates, $\ket{\nu \vb{q}} = Q_{\nu\vb{q}}^\dagger \ket{0}$. The TDHF equations exhibit an intrinsic particle-hole symmetry that, combined with inversion symmetry, ensures that the resulting spectrum is symmetric~\cite{kwanMoireFractionalChern2025,roweEquationsofMotionMethodExtended1968,NakadaPhysical2016}, with every $\omega_{\nu \vb{q}}>0$ possessing a partner $-\omega_{\nu \vb{q}}<0.$ The positive half of the spectrum, which we label $\omega_{1\vb{q}}, \omega_{2\vb{q}}, \ldots, \omega_{M_q\vb{q}},$ corresponds to the creation operators $Q^\dagger_{\nu\vb{q}}$, while the negative solutions correspond to the conjugate annihilation processes.

It can be shown that solving the TDHF equations~\eqref{eq:TDHF_matrix} is equivalent to diagonalizing the following bosonic Hamiltonian~\cite{wolfQuasibosonApproximationYields2024a},
\begin{equation}
    \label{eq:HB2}
    \mathcal{H}_B = E_{\mathrm{HF}} + \frac{1}{2}\sum_{\vb{q}} \left[ \begin{bmatrix}
        b_{\vb{q}}^\dagger & b_{-\vb{q}}^T
    \end{bmatrix} 
    S_{\vb{q}} 
    \begin{bmatrix}
        b^{\phantom{\dagger}}_{\vb{q}} \\ (b_{-\vb{q}}^\dagger)^T
    \end{bmatrix} - \tr A_{\vb{q}} \right]
    ,
\end{equation}
where we have defined the basis $b_{\vb{q}}^\dagger = \left[b^\dagger_{\vb{q}}(\vb{k}_1), b^\dagger_{\vb{q}}(\vb{k}_2), \ldots, b^\dagger_{\vb{q}}(\vb{k}_{M_q}) \right]$ for $\vb{k}_i\in K_q$. The TDHF amplitudes $X_{\nu\vb{q}}$ and $Y_{\nu \vb{q}}$ define a bosonic Bogoliubov transformation that diagonalizes $\mathcal{H}_B$, which then takes the natural form of a sum of harmonic oscillators,
\begin{equation}
    \label{eq:HB3}
    \mathcal{H}_B = E_{\mathrm{HF}} + \frac{1}{2}\sum_{\vb{q}} \sum_{\nu =1}^{M_q} \left[\omega_{\nu \vb{q}} ( Q^\dagger_{\nu \vb{q}} Q^{\phantom{\dagger}}_{\nu \vb{q}} + 1) - \tr A_{\vb{q}}\right]
    .
\end{equation}
Recasting the Hamiltonian in this form makes it clear that the correlation energy of the system is the difference between the zero-point energy of the collective modes and the bare energy of the elementary particle-hole excitations, captured by the trace of $\hat{A}_{\vb{q}}$:
\begin{align}
    \label{eq:corr}
    \varepsilon_\mathrm{corr} = \frac{1}{2N}\sum_{\vb{q}} \left[ \sum_{\nu = 1}^{M_q} \omega_{\nu\vb{q}} - \tr A_{\vb{q}}\right]
    .
\end{align}

The structure of $S_{\vb{q}}$ also encodes the energetic stability of the HF ground state. Consider a Slater determinant that is slightly perturbed away from the HF ground state~\cite{THOULESS1960225, ge2026nematicwignercrystalsrhombohedral},
\begin{align}
    \ket{\alpha_{\vb{q}}} = \exp\left[b_{\vb{q}}^\dagger \alpha_{\vb{q}} - \alpha^\dagger_{\vb{q}} b^{\phantom{\dagger}}_{\vb{q}} + b_{-\vb{q}}^\dagger \alpha_{-\vb{q}}  - \alpha_{-\vb{q}}^\dagger b^{\phantom{\dagger}}_{-\vb{q}}\right] \ket{\mathrm{HF}}
    ,
\end{align}
where $\alpha_{\vb{q}}^T := \left[
    \alpha^{\phantom{\dagger}}_{\vb{q}}(\vb{k}_1), \alpha^{\phantom{\dagger}}_{\vb{q}}(\vb{k}_2), \ldots,\alpha^{\phantom{\dagger}}_{\vb{q}}(\vb{k}_{M_q})
\right]$, with $\vb{k}_i \in K_q$, and $|\alpha^{\phantom{\dagger}}_{\vb{q}}| \ll 1$. It is straightforward to show that the energy of this state is, to quadratic order in $\alpha^{\phantom{\dagger}}_{\vb{q}}$,
\begin{equation}
\begin{aligned}
    E(\alpha^{\phantom{\dagger}}_{\vb{q}}) - E_{\mathrm{HF}} &= \mel{ \alpha^{\phantom{\dagger}}_{\vb{q}}}{\mathcal{H}}{\alpha^{\phantom{\dagger}}_{\vb{q}}} - \mel{\mathrm{HF}}{\mathcal{H}}{\mathrm{HF}} \\
    &\approx
    \begin{bmatrix}
        \alpha_{\vb{q}}^\dagger & \alpha_{-\vb{q}}^T
    \end{bmatrix}
    S_{\vb{q}}
    \begin{bmatrix}
        \alpha^{\phantom{\dagger}}_{\vb{q}} \\
        \alpha_{-\vb{q}}^*
    \end{bmatrix}.
\end{aligned}    
\end{equation}
If $S_{\vb{q}}$ is positive definite, then any perturbation away from $\ket{\text{HF}}$ will increase the energy, and $E_{\text{HF}}$ is a genuine local minimum in the space of Slater determinants. In contrast, negative eigenvalues of $S_{\vb{q}}$ indicate that the HF energy can be lowered by perturbing away from $\ket{\text{HF}}$, rendering the reference state unstable.

This situation manifests in the spectrum of the TDHF matrix, $\Sigma S_{\vb{q}},$ as imaginary or complex eigenvalues, as well as modes with non-normalizable eigenvectors~\cite{NakadaPhysical2016}. Explicitly, if we diagonalize according to
\begin{align}
    S_{\vb{q}} 
\begin{bmatrix}
    u_{\vb{q}} \\
    u_{-\vb{q}}
\end{bmatrix} = \xi_{\vb{q}}
\begin{bmatrix}
    u_{\vb{q}} \\
    u_{-\vb{q}}
\end{bmatrix}
,
\end{align}
the presence of a negative eigenvalue, $\xi_{\vb{q}}< 0,$ means that the HF energy can be lowered by taking a linear combination of $\ket{\mathrm{HF}}$ with the Slater determinant characterized by the vector $u_{\pm\vb{q}}$. Although such instabilities limit the accuracy of TDHF when they occur, knowledge of such instabilities can provide important insight into the underlying physics.

The collective mode spectrum also determines the response functions of the system. By constructing the particle-hole Green's function from the TDHF eigensystem, one can show that the response function between two operators $\hat{\mathcal{O}}_{1\vb{q}}$ and $\hat{\mathcal{O}}_{2\vb{q}}$ takes the form
\begin{widetext}
\begin{equation}
    \chi_{\mathcal{O}_1 \mathcal{O}_2}(\vb{q},\omega) = \frac{1}{A} \sum_{\nu=1}^{M_q}\left(\frac{\bra{0} \hat{\mathcal{O}}_{1,-\vb{q}}\ket{\nu \vb{q}}\bra{\nu \vb{q}} \hat{\mathcal{O}}_{2\vb{q}} \ket{0}}{\omega - \omega_{\nu \vb{q}} + i0^+} 
    - 
    \frac{\bra{0} \hat{\mathcal{O}}_{2,-\vb{q}} \ket{\nu \vb{q}} \bra{\nu \vb{q}} \hat{\mathcal{O}}_{1\vb{q}} \ket{0} }{\omega + \omega_{\nu \vb{q}} + i0^+} \right)
    .
\end{equation}
\end{widetext}
In this work, we are interested in density and current responses. The band-projected density operator has matrix elements 
\begin{equation}
\label{eq:rho_me}
\begin{aligned}
    \mel{\nu \vb{q}}{\hat{\rho}_{\vb{q}}}{0} = &\sum_{\vb{k} \in K_q} \mathcal{F}(\vb{k}+\vb{q},\vb{k}) X^*_{\nu\vb{q}}(\vb{k})
    \\
    + &\sum_{\vb{k}\in K_{-q}} \mathcal{F}(\vb{k},\vb{k}-\vb{q})Y^*_{\nu\vb{q}}(\vb{k})
    ,
\end{aligned}            
\end{equation}
which can be further resolved into pseudospin components. The form factor is defined as 
$\mathcal{F}(\vb{k},\vb{k}') = \sum_\sigma u^*_\sigma(\vb{k})u_\sigma(\vb{k}')$, where $\sigma$ labels the pseudospin components of the pseudospinor, so the pseudospin resolved band-projected density operator is obtained simply by restricting the sum over $\sigma$ to a particle component:
\begin{equation}
    \hat{\rho}_{\vb{q} \sigma} = \sum_{\vb{k}} \mathcal{F}_{\sigma}(\vb{k}+\vb{q},\vb{k}) \Cdag{\vb{k} + \vb{q}} \C{\vb{k}},
\end{equation}
where we have defined
\begin{align}
    \mathcal{F}_\up(\vb{k},\vb{k}') &= \frac{1}{\sqrt{(1+\lambda^2k^2)(1+\lambda^2k'^{2})}} \\
    \mathcal{F}_\dn(\vb{k},\vb{k}') &= \frac{\lambda^2(k_x + ik_y)(k_x'  -ik_y')}{\sqrt{(1+\lambda^2k^2)(1+\lambda^2k'^{2})}}
    .
\end{align}
The orbital resolved response function is $\chi_{\sigma \sigma'}:= \chi_{\rho_\sigma \rho_{\sigma'}}$, and the relevant matrix elements $\bra{\nu \vb{q}} \hat{\rho}_{\vb{q}\sigma} \ket{0}$ are analogous to (\ref{eq:rho_me}), with the appropriate form factor.

The current-current response takes a similar form, but the relevant matrix elements are those of the band-projected current operator~\cite{bruusManyBodyQuantumTheory2004},
\begin{align}
    \hat{\vb{j}}_{\vb{q}} = \frac{\hbar}{m}\sum_{\vb{k}}\left(\vb{k}-\frac{\vb{q}}{2}\right)\mathcal{F}(\vb{k}-\vb{q},\vb{k})c^\dagger_{\vb{k}-\vb{q}}\C{\vb{k}}.
\end{align}
given by
\begin{equation}
\begin{aligned}
    \mel{\nu \vb{q}}{\hat{\vb{j}}_{-\vb{q}}}{0}
    &=
    \frac{\hbar}{m} \left[ \sum_{\vb{k} \in K_q} \left( \vb{k} + \frac{\vb{q}}{2} \right) \mathcal{F}(\vb{k}+ \vb{q}, \vb{k}) X^*_{\nu \vb{q}}(\vb{k}) \right. \\ 
    &\left.+\sum_{\vb{k} \in K_{-q}}\left(\vb{k} - \frac{\vb{q}}{2} \right)\mathcal{F}(\vb{k},\vb{k} - \vb{q}) Y^*_{\nu \vb{q}} (\vb{k}) \right]
    .
\end{aligned}
\end{equation}
We decompose the current operator into its longitudinal and transverse components, defined as $\hat{\vb{j}}_{\vb{q},\parallel} = \frac{\hat{\vb{j}}_{\vb{q}}\cdot\vb{q}}{q^2} \vb{q}$ and $\hat{\vb{j}}_{\vb{q},\perp} = \hat{\vb{j}}_{\vb{q}}-\hat{\vb{j}}_{\vb{q},\parallel}$, respectively. 
The longitudinal current-current response function is entirely determined by the density-density response through the continuity relation $\omega^2 \chi_{\rho\rho}(\vb{q},\omega) = q^2\chi_{j_\parallel j_\parallel}(\vb{q},\omega)$, so we focus on the transverse current response,  $\chi_{j_\perp j_\perp}$.

When calculating the FL correlation energy, we use a momentum discretization of $dk = 0.03 k_\mathrm{F}$ and a momentum cutoff of $q<5k_\mathrm{F}$, and when calculating the FL response functions we set $dk = 0.08 k_\mathrm{F}$. This formulation of TDHF assumes that translation symmetry is preserved. To modify the approach for the crystalline states, one allows for translation symmetry breaking at the Hartree-Fock level and constructs particle-hole excitations on top of the symmetry broken state accordingly. The details of the implementation of TDHF for crystalline systems is well-documented in the literature~\cite{jainElementaryExcitationsMelting2025, huoDoesMoireMatter, kwanMoireFractionalChern2025,  dongPhononsElectronCrystals2025, desrochersElasticResponseInstabilities2026, hirsbrunnerTopologicalPhononsAnomalous2026a,miao2026retardedinteractionoppositechiral, ge2026nematicwignercrystalsrhombohedral}. To obtain the HF ground state, we converge the density matrix using the periodic Pulay mixing~\cite{rohwedder2011analysis}, and stop the iteration when self-consistency is achieved to a tolerance of $1\times10^{-8}.$ When computing the correlation energy, we employ an $18\times18$ lattice and include six momentum shells. When computing collective modes and response functions along high-symmetry lines, we use a $30\times30$ lattice and include four momentum shells.

\begin{figure}
    \centering    \includegraphics[width=\linewidth]{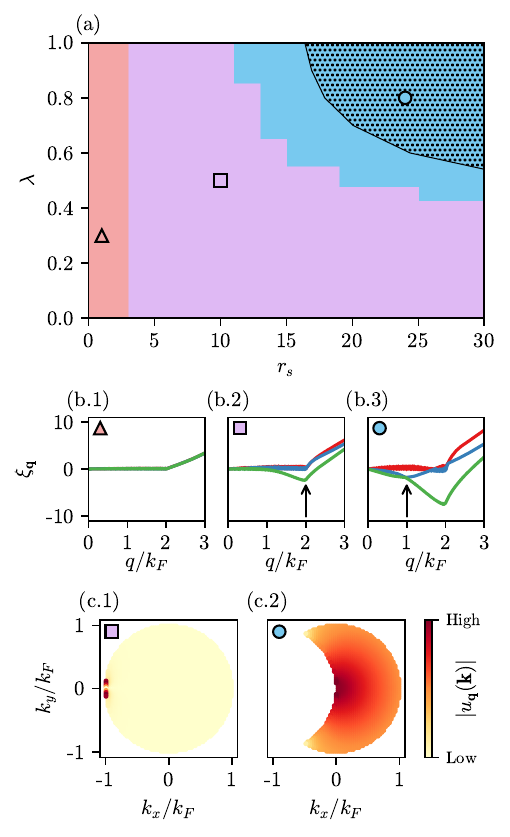}
    \caption{(a) The liquid--crystal phase diagram. The dotted region indicates where the crystal phase is lower in energy than the liquid, with the energies computed via TDHF. The colour indicates the nature of any instabilities of the mean-field FL ground state. In the red region, no eigenvalues of the stability matrix $S_{\vb{q}}$ are significantly negative. In the purple region, $S_{\vb{q}}$ possesses one band with negative eigenvalues, peaked at $q=2k_{\mathrm{F}}$. In the blue region, there are two bands with negative eigenvalues, one peaked at $q=2k_{\mathrm{F}}$ and one at $q=k_{\mathrm{F}}$. (b) Band structures of $S_{\vb{q}}$ at three representative points in the phase diagram. Each plot depicts the three lowest eigenvalues of $S_{\vb{q}}$ as a function of $q$ at the parameter points indicated by the symbols in the top left. Different colours are used to differentiate distinct bands. (c) The magnitude of eigenvectors as a function of $\vb{k}$ at parameter points indicated by the symbols in the upper left, corresponding to the eigenvalues marked by arrows in (b.2) and (b.3).}
    \label{fig:pd}
\end{figure}

\section{Phase diagram}
\label{sec:pd}
We first apply TDHF to understand how the presence of quantum geometry reshapes the liquid--crystal phase boundary. To begin, we determine the location of the phase boundary by comparing the ground state energies of the liquid and crystal phases. The ground state energy of the liquid phase is simply the sum of the HF energy and the correlation energy contributed by the collective modes~(\ref{eq:corr}). In the crystal phase, we also include the standard Madelung energy correction to account for the long-range Coulomb interaction and minimize finite-size effects~\cite{PARRY1975433,kawataRapid2001,soejimaJelliumModelAnomalous2025}. The resulting phase diagram is plotted in Fig.~\ref{fig:pd}~(a), in which the dotted region denotes where the crystalline state is lower in energy. In the case of trivial form factors, $\lambda = 0$, our calculation finds that the Fermi liquid is always lower in energy, with no transition to a crystal for $r_s\leq 30$. This is starkly different from the mean field case, which predicts the $\lambda=0$ transition at $r_s \sim2$. Correlation energy corrections generally act to favour the FL ground state, with recent QMC results placing the transition for the conventional polarized two-dimensional electron gas at $r_s \sim 30$, at the very edge of our phase diagram~\cite{azadiQuantumMonteCarlo2024}. As TDHF is a perturbative technique, it is not constrained by the variational principle, and we find that the ground state energy of the FL within TDHF is more negative than the QMC result, pushing the phase boundary to even higher $r_s$. We note that this effect is less significant in spinful FLs, wherein spin fluctuations enhance the screening of the Coulomb interaction, reducing correlations and allowing TDHF to produce energies competitive with QMC~\cite{wolfQuasibosonApproximationYields2024a}. Introducing finite $\lambda$ dramatically lowers the critical $r_s$, with a phase boundary forming at $r_s \sim 30$ for $\lambda \sim 0.5$ and curves towards $r_s\sim 16$ as $\lambda$ is increased to 1. The presence of quantum geometry thus has a large influence on the boundary, with finite $\lambda$ driving the system to crystallize at higher densities.

Our findings qualitatively agree with those previously obtained via an NQS variational Monte Carlo approach~\cite{valentiQuantumGeometryDriven2025}, which placed the phase boundary around $r_s \sim 28$ at $\lambda=0$, $r_s \sim 12$ at $\lambda\sim0.4$, and $r_s = 8 $ at $\lambda \sim 1$. As in the $\lambda=0$ case, we see that TDHF has a tendency to produce a correlation energy for the FL that is more negative than the exact value, which pushes the phase boundary to larger values of $r_s$ and results in a quantitative disagreement between the NQS and TDHF phase boundaries. The accuracy of TDHF also suffers when the HF ground state is not a local minimum of the energy landscape. Such an instability manifests as imaginary and complex eigenvalues in the solution of the TDHF equations~\cite{THOULESS1960225, NakadaPhysical2016}, which complicates the interpretation of the obtained correlation energy. For large values of $r_s$, we indeed find complex eigenvalues in the TDHF spectrum, indicating an instability in the HF ground state. While the above considerations limit the quantitative accuracy of the TDHF liquid--crystal phase boundary, its qualitative agreement with the NQS phase boundary strengthens the claim that quantum geometry can shift the crystallization transition to higher densities.

Despite these limitations, TDHF provides additional insight into the role of quantum geometry in electron crystallization through the stability matrix $S_{\vb{q}}$. Here we study the eigenvalues and the eigenvectors of the stability matrix across the phase diagram to gain additional information about the physics driving the crystallization transition in the $\lambda-$jellium model. We seek to identify the portions of the phase diagram that host negative eigenvalues of $S_{\vb{q}},$ which indicate an energetic instability of the HF ground state, and characterize at which momenta those negative eigenvalues occur.

Three distinct regions emerge, as shown in Fig.~\ref{fig:pd}~(a). We first consider the region at small $r_s$ plotted in red, in which $S_{\vb{q}}$ possesses no eigenvalues that are significantly negative, indicating a stable HF ground state. In Fig.~\ref{fig:pd}~(b.1) we plot the lowest three eigenvalues of $S_{\vb{q}}$ as a function of $q$ at a representative point marked by the triangle in the phase diagram. The eigenvalues are nearly degenerate and collapse into a single line. For $0<q<2k_\mathrm{F}$, the eigenvalues are nearly zero and correspond to gapless excitations across the Fermi surface. In contrast, the eigenvalues are strictly positive  for $q>2k_\mathrm{F}$, as there are no gapless excitations available and any particle-hole excitation with momentum $q>2k_\mathrm{F}$ will raise the energy.

\begin{figure*}
    \centering    \includegraphics[width=\linewidth]{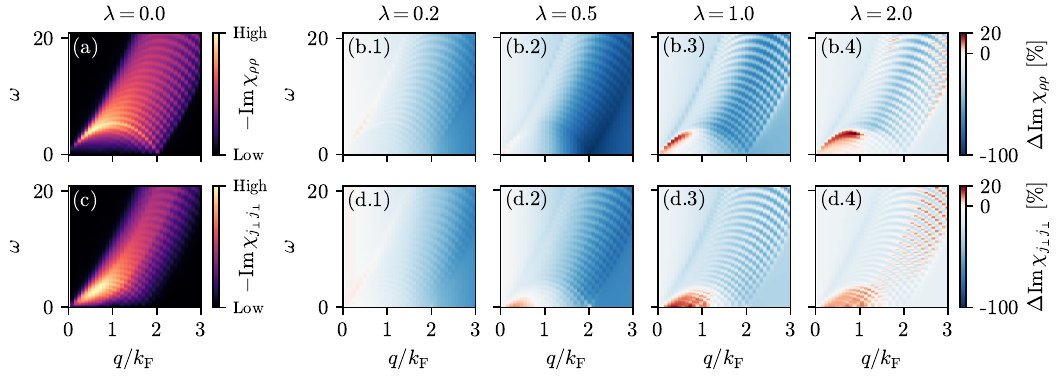}
    \caption{(a) The imaginary part of the density-density response, $\chi_{\rho \rho}(\vb{q},\omega),$ in the Fermi liquid phase at $r_s=1$ and $\lambda=0$. (b) The percent change in the imaginary part of the density-density response compared to the $\lambda=0$ case for four values of $\lambda$, demonstrating how non-trivial quantum geometry reshapes the response. (c) The imaginary part of the transverse current-current response, $\chi_{j_\perp j_\perp}(\vb{q},\omega),$ for $r_s=1$ and $\lambda=0$. (d) The percent change in the imaginary part of the transverse current-current response compared to the $\lambda=0$ case.
    }
    \label{fig:response}
\end{figure*}

In Fig.~\ref{fig:pd}~(b.2) we plot the lowest three eigenvalues of $S_{\vb{q}}$ at another representative point at higher $r_s$ and $\lambda,$ marked by the square. Here the lowest band is negative over a wide range of momenta, obtaining a minimum at $q=2k_{\mathrm{F}},$ while the higher bands remain positive. At even higher $r_s$ and $\lambda$, at the point indicated by the circle marker in the phase diagram, we see that the lowest two bands obtain negative values, as plotted in Fig.~\ref{fig:pd}~(b.3). The most negative eigenvalue remains pinned at $q=2k_{\mathrm{F}},$ but the minimum of the second band occurs at $q=k_{\mathrm{F}}.$ Motivated by these observations, we classify the remainder of the phase diagram according to the number of bands of $S_{\vb{q}}$ with negative eigenvalues. In the purple region, the stability matrix exhibits a single negative band that is always peaked at $q=2k_{\mathrm{F}},$ whereas $S_{\vb{q}}$ has two negative bands, one each peaked at $q=k_{\mathrm{F}}$ and $q=2k_{\mathrm{F}},$ in the blue region.

The nature of the instabilities in the purple and blue regions are revealed by examining the eigenvectors at the peaks of the negative bands. We first consider the $q=2k_{\mathrm{F}}$ instability that is present in both regions, plotting in Fig.~\ref{fig:pd}~(c.1) the eigenvector corresponding to the eigenvalue indicated by the arrow in Fig.~\ref{fig:pd}~(b.2). We take $\vb{q}$ to point in the $+\vb{x}$ direction, without loss of generality as the system is isotropic, and examine the magnitude of the components $|u_{\vb{q}}(\vb{k})|$. At $q=2k_\mathrm{F}$, the domain of the eigenvector, $K_q,$ is simply the full Fermi sea. The eigenvector is peaked at the left edge of the Fermi sea, around $\vb{k}=(-k_{\mathrm{F}}, 0),$ and vanishes elsewhere. This structure indicates that the ground state energy can be significantly lowered by exciting electrons from one edge of the Fermi surface to the other, corresponding to the well-known $2k_\mathrm{F}$ charge density wave instability of the Fermi gas~\cite{overhauserSpinDensityWaves1962,overhauserChargedensityWavesIsotropic1978,colonnaCorrelationEnergyExactexchange2014,wolfQuasibosonApproximationYields2024a}. This instability is a consequence of a sharp Fermi surface, so it is independent of the quantum geometry and persists even to large values of $\lambda$.

On the other hand, the instability at $q=k_{\mathrm{F}}$ that characterizes the blue region is a consequence of quantum geometry, only appearing for large values of $\lambda.$ In Fig.~\ref{fig:pd}~(c.2), we plot a representative eigenvector of this second instability, at the point marked by the arrow in Fig.~\ref{fig:pd}~(b.3)~\footnote{There is an avoided crossing between the lowest two bands at $q$ slightly below $q=k_{\mathrm{F}},$ across which the qualitative natures of the eigenvectors of the bands exchange. In Fig.~\ref{fig:pd}~(c.2) we take $q=k_{\mathrm{F}}$ and consider the second band, selecting the eigenvector distinct from that characteristic of the $q=2k_{\mathrm{F}}$ instability.}. We find that the eigenvector $|u_{\vb{q}}(\vb{k})|$ has a wide peak centred at $\vb{k}=0,$ but remains finite over the entirety of the crescent-shaped domain $K_q$. This indicates that the HF ground state energy can be reduced by shifting occupation from small values of $\vb{k}$ to larger ones, with the largest change caused by ejecting electrons from the $\Gamma$ point out to the Fermi surface. The distribution of this eigenfunction is a natural consequence of the energetic penalty arising from the exchange term being reduced around the $\Gamma$ point, where the quantum geometry is strongest. One may question if this instability is towards the AFL, the formation of which is driven by similar physics. This is not the case, as the HF ground state energy of the FL is lower than the AFL for all $\lambda<1$~\cite{soejimaJelliumModelAnomalous2025}. Regardless, we consider the instability from FL to AFL at larger values of $\lambda$ in Appendix~\ref{sec:AFLap}. We note that this second instability at $q=k_{\mathrm{F}}$ emerges shortly before the liquid--crystal transition in the phase diagram. The coincidence of these two boundaries strongly hints that the fluctuations driving the instability play a major role in the competition between the liquid and crystal phases.

\section{Fermi liquid response functions}

\begin{figure}
        \centering
        \includegraphics[width=\linewidth]{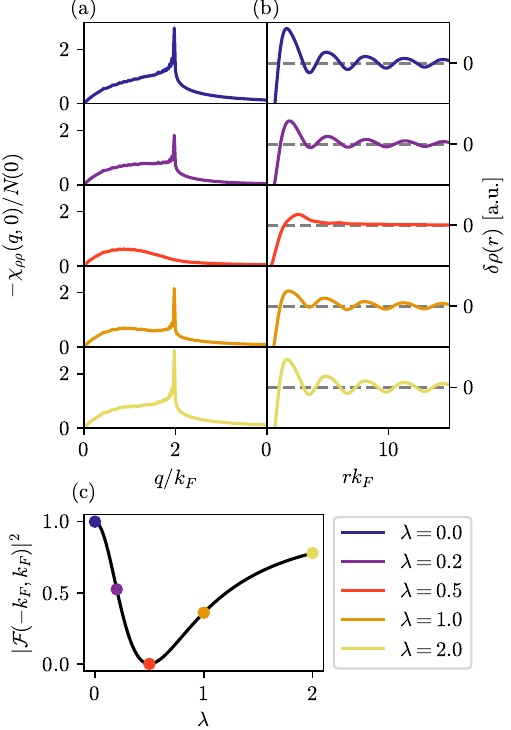}
        \caption{(a) The static limit of the density-density response $\chi_{\rho\rho}(q,\omega = 0)$ in the Fermi liquid at $r_s=1$ for a range of different $\lambda$. The sharp peaks at $q=2k_\mathrm{F}$ correspond to Friedel oscillations in the real space charge density. (b) The induced real space charge density $\delta \rho(r)$, given by a Fourier transform of $\chi_{\rho\rho}(q,0)$. The peak at $2k_\mathrm{F}$ and corresponding real space oscillations vanish at $\lambda=0.5$. (c) The form factor $|\mathcal{F}(-k_\mathrm{F},k_\mathrm{F})|^2$, which connects states at opposite edges of the Fermi surface, the relevant scattering states for Friedel oscillations. The matrix element is completely suppressed at $\lambda=0.5$.
        }
        \label{fig:friedel}
\end{figure}

We now turn to consider the role that quantum geometry plays in determining the properties of the Fermi liquid phase far from the crystallization transition, computing the density-density and current-current responses within TDHF. In Fig.~\ref{fig:response}~(a) we plot the imaginary part of the density-density response of the conventional FL at $r_s=1,$ $\lambda=0$. The spectral weight contains a peak near $(q=0,\omega=0)$, corresponding to the plasmon mode, as well as a peak arcing towards $(q=2k_{\mathrm{F}},\omega=0)$. Similarly to the FL phase of the spinful 2D electron gas, the plasmon is fully embedded in the particle-hole continuum at this value of $r_s$~\cite{wolfQuasibosonApproximationYields2024a}. At small $q$, the spectral response is exhausted by the plasmon mode, whereas the response spreads out over the entire particle-hole continuum for larger $q$. As $r_s$ is increased, the peak at $(\omega=0,q=2k_\mathrm{F})$ strengthens, eventually becoming the charge density wave instability discussed in the previous section. However, the FL is still stable for $r_s = 1$.

In Fig.~\ref{fig:response}~(b.1-b.4) we plot the change in the density-density response as a function of $\lambda$, relative to the $\lambda=0$ case, i.e. $\Delta\operatorname{Im} \chi_{\rho\rho} = (\operatorname{Im} \chi_{\rho\rho}^\lambda-\operatorname{Im} \chi_{\rho\rho}^{\lambda=0}) /\operatorname{Im} \chi_{\rho\rho}^{\lambda=0}$. As $\lambda$ is turned on, the primary impact of the quantum geometry is the suppression of the response on the right edge of the particle-hole continuum. This suppression reaches a maximum at $\lambda = 1/2,$ and spectral weight returns for larger values of $\lambda$. Physically, the right edge of the particle-hole continuum corresponds to scattering events that traverse the entire width of the Fermi sea. The relevant form factor connecting these states becomes small for certain values of $\lambda$, causing the suppression of the response. The suppression is particularly pronounced in the static limit $(\omega \to 0)$ at $q = 2k_\mathrm{F}.$

Investigating this effect further, we specialize to the static limit of the density-density response, $\chi_{\rho\rho}(\vb{q},\omega = 0)$. In Fig. \ref{fig:friedel}~(a), we plot the static response for the same five values of $\lambda$ as in Fig.~\ref{fig:response}. In general, the induced density is given to linear order by $\delta \rho(\vb{q},\omega) = \chi_{\rho\rho}(\vb{q},\omega) \varphi^\mathrm{ext}(\vb{q},\omega)$, where $\varphi^\mathrm{ext}(\vb{q},\omega)$ is an external potential with wavevector $\vb{q}$ and frequency $\omega$. As such, $\chi_{\rho\rho}(\vb{q},0)$ can be understood as density induced by a static, $\delta$-function impurity in real space. Examining $\chi_{\rho\rho}(\vb{q},0)$ for $\lambda = 0$, we see that there is a large peak at $q = 2k_\mathrm{F}$, indicating that the density surrounding an impurity oscillates with a characteristic wavevector of $2k_\mathrm{F}$. These are the well-established Friedel oscillations of an electron gas~\cite{friedelMetallicAlloys1958}. As $\lambda$ is increased, the suppression of the response shrinks the magnitude of the $2k_\mathrm{F}$ peak, which vanishes completely at $\lambda = 1/2$, before reemerging as $\lambda$ is further increased. 
At $\omega = 0, q=2k_\mathrm{F}$, the relevant excitations are backscattering events, in which an electron scatters from one side of the Fermi surface to the other, i.e. from $\vb{k}$ to $-\vb{k}$, with $|\vb{k}| = k_\mathrm{F}$. Since the system is isotropic, the magnitude of the relevant form factor for such backscattering events does not depend on the momentum direction, and is given by $|\mathcal{F}(-k_\mathrm{F},k_\mathrm{F})|^2 = (1-\lambda^2 k_\mathrm{F}^2)^2/(1 + \lambda^2 k_\mathrm{F}^2)^2$. The form factor vanishes when $\lambda = k_\mathrm{F}^{-1} = 1/2$, suppressing the Friedel oscillation peak at $q = 2k_\mathrm{F}$. 

The suppression of backscattering seen here is similar in spirit to the suppression of Friedel oscillations in graphene, where backscattering is forbidden by the chiral nature of the electrons~\cite{PhysRevB.93.035413}, as well as the observation of damped Friedel oscillations in topological insulator surface states arising from spin-momentum locking~\cite{PhysRevB.89.195417,PhysRevLett.104.016401}. Additionally, a recent work studied a related phenomenon in which systems with ``hotspots'' of quantum geometry induce novel quantum geometric Friedel oscillations~\cite{ma2026quantumgeometricfriedeloscillations}.

Returning to the density-density response in Fig.~\ref{fig:response}, the other notable feature is that the dispersion of the plasmon is pushed to lower frequency as $\lambda$ is increased. This is particularly visible in plots (b.3) and (b.4), where the red region reveals that the peak of the plasmon mode has shifted to lower energy at small $q$. The origin of this behaviour can be understood through a random-phase approximation (RPA) analysis of the plasmon dispersion in the small $q$ limit. While this approximation neglects the exchange terms that are included in TDHF calculations, its simplicity provides valuable analytical insight into why the energy of the plasmon mode is reduced. Performing an expansion of the RPA susceptibility for $q\ll k_\mathrm{F}$, the lowest-order correction to the plasmon mode for finite $\lambda$ is given by
\begin{align}
        \omega_\mathrm{p}^\mathrm{RPA,\lambda} - \omega_\mathrm{p}^\mathrm{RPA,0} \sim - \frac{\omega^0_\mathrm{p}q^2}{4\pi k_\mathrm{F}^2}\int_{k < k_\mathrm{F}}d^2 \vb{k} \tr g(\vb{k})
        ,
\end{align}
as derived in Appendix~\ref{sec:RPAap}. Here $\omega_\mathrm{p}^\mathrm{RPA,\lambda}$ is the RPA plasmon dispersion at finite $\lambda$, $\omega_\mathrm{p}^\mathrm{RPA,0}$ is the RPA plasmon dispersion at $\lambda=0$, and $\omega_\mathrm{p}^0 = \frac{2 \sqrt{2 q a_B}}{r_S}~\mathrm{Ry}$ is the classical two-dimensional plasmon dispersion. The quantity $g(\vb{k})$ is the quantum metric, defined as $g_{ab}(\vb{k}) = \operatorname{Re} \mel{\partial_{k_a} u_{\vb{k}}}{\left( \mathds{1} - \ket{u_{\vb{k}}} \bra{u_{\vb{k}}} \right)}{ \partial_{k_b} u_{\vb{k}}}$, which measures the orthogonality of nearby states~\cite{provostRiemannianStructureManifolds1980}. Interestingly, we see that there is a quantum geometric correction to the plasmon frequency, given by the trace of the quantum metric integrated over the Fermi sea. As the metric is positive semi-definite, this correction always acts to reduce the plasmon dispersion in systems with non-trivial quantum geometry. This reduction of the plasmon frequency is consistent with results on the Wigner crystal side, where finite Berry curvature was found to reduce the speed of the longitudinal phonon - the crystalline analog of the plasmon~\cite{dongPhononsElectronCrystals2025}. 

While the density-density response has a well-defined plasmon, the transverse current-current response $\chi_{j_\perp j_\perp}$, plotted on the second row of Fig.~\ref{fig:response}, has no such sharp collective mode. Instead, the $\lambda = 0$ response, shown in Fig.~\ref{fig:response}(c), is broadly distributed over the particle-hole continuum. As for the density response, the spectral weight on the right edge of the particle-hole continuum is suppressed for small values of $\lambda$ and reappears as $\lambda$ is tuned above $\lambda=0.5$. However, the transverse current response is already suppressed on the right edge of the particle-hole continuum in the $\lambda=0$ limit because the states involved always carry only momentum parallel to the scattering vector. As such, the impact of quantum geometry here is negligible. In contrast, the broad peak at small $q$ is pushed down as $\lambda$ increases, as evidenced by the red regions in Fig.~\ref{fig:response}(d.1-d.4), similar to the dampening of the plasmon dispersion. 

To gain a better physical picture of why the spectral weight redistributes as $\lambda$ is tuned, we also consider the orbital-resolved density response. We construct the orbital-resolved density response tensor,
\begin{align}
        \label{eq:OR_matrix}
        \hat\chi = \begin{pmatrix}
            \chi_{\up\up} & \chi_{\up\dn} \\
            \chi_{\dn\up} & \chi_{\dn\dn}
        \end{pmatrix}
        ,
\end{align}
and diagonalize the anti-Hermitian part of it, $\hat\chi_{\mathrm{AH}} = \frac{1}{2i}(\hat\chi - \hat\chi^\dagger),$ to identify the eigenmodes of the orbital-resolved density response. The eigenvalues $\chi_{+}$ and $\chi_{-},$ defined such that $|\chi_{+}|>|\chi_-|,$ give the spectral weight of the normal modes, and the corresponding eigenvectors reveal the orbital character of the response. In general, we find that there is one dominant eigenmode $\chi_{+}$, so we focus our analysis on this mode. We write the eigenvector of the dominant $\chi_{+}$ mode as $(\cos \theta/2, \sin \theta/2 e^{i\phi}),$ such that $\theta$ determines the weight of the response on each orbital and $\phi$ gives the relative phase of the response between the orbitals. 

\begin{figure}
    \centering
    \includegraphics[width=\linewidth]{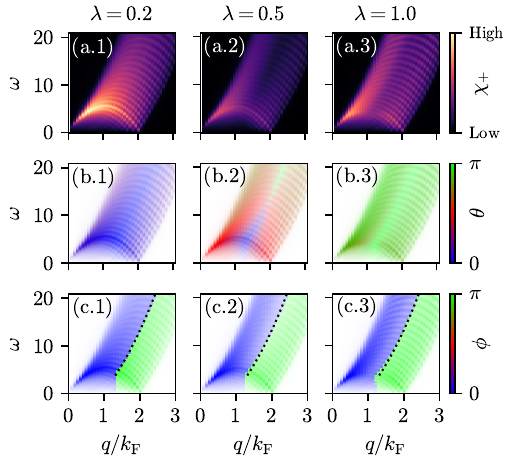}
    \caption{(a.1-a.3) The dominant eigenvector of the orbital-resolved density-density response, $\chi_+$, in the Fermi liquid at $r_s=1$, plotted for $\lambda=0.2,$ $0.5,$ and $1.0,$ respectively. (b.1-b.3) The angle $\theta$ that characterizes the orbital polarization of the response for the same values of $\lambda$. When $\theta = 0$, the response is entirely polarized on the pseudospin $\up$ orbital, and entirely polarized on the $\dn$ orbital when $\theta=\pi$. (c.1-c.3) The angle $\phi$ that characterizes the phase of the inter-orbital response. This phase is only meaningful if $0<\theta<\pi$, in regions where both orbitals contribute. $\phi=0$ indicates an in-phase response, while $\phi=\pi$ indicates an out-of-phase response. The black dotted lines depict a perturbative calculation of the location where $\phi$ switches from $0$ to $\pi$, given by Eq.~\eqref{eq:phase_switch}.}
    \label{fig:OR_response}
\end{figure}

In the first column of Fig.~\ref{fig:OR_response}, we plot $\chi_+,$ $\theta,$ and $\phi$ for the orbital-resolved density-density response of the Fermi liquid at $r_s = 1$ and $\lambda=0.2.$ The value of $\theta$ deviates only slightly from zero, indicating that the response is strongly pseudospin-polarized on the $\up$ orbital for this small value of $\lambda.$ As such, $\chi_+$ closely matches the total density response plotted in Fig.~\ref{fig:response}. 

\begin{figure}
    \centering
    \includegraphics[width=\linewidth]{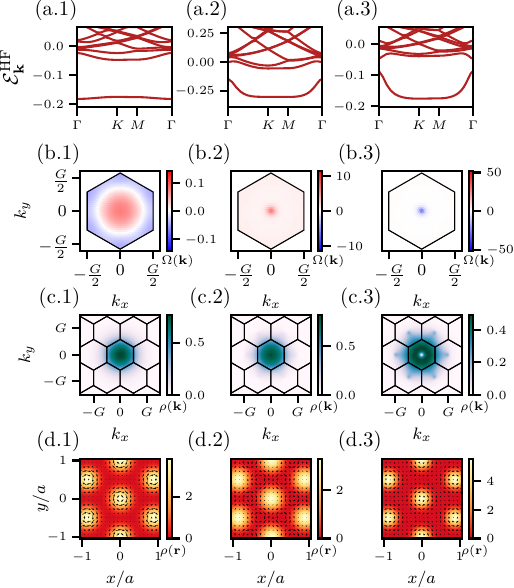}
    \caption{The (a) HF band structure plotted along high-symmetry lines across the BZ, (b) Berry curvature of the lowest occupied band, (c) momentum space occupation $\rho(\vb{k})$, and (d) real-space charge density, of three different crystalline phases. Panels (a.1-d.1) correspond to the WC at $r_s = 20,\lambda = 0.2,$ (a.2-d.2) correspond to the AHC at $r_s=10, \lambda=1.6,$ and (a.3-d.3) correspond to the HWC at $r_s=20, \lambda=2.0$. The charge density in (d) is depicted by the coloured map, and the ground state current density is overlaid with arrows.}
    \label{fig:HF_crystal}
\end{figure}

Examining the relative phase, $\phi$, of the response between the orbitals, we see that it abruptly jumps from $\phi=0$ to $\phi=\pi$ on a line traversing the centre of the particle-hole continuum. This behaviour originates from the in-plane component of the pseudospinor skyrmion texture winding as $\vb{k}$ rotates around the origin. As a result of this winding, scattering processes acquire a relative phase determined by the angle between initial and final momenta. At small $q$, the density operator connects states with nearly parallel momenta and closely aligned pseudospin orientations, producing the observed in-phase response. In contrast, in the large-$q$ limit, the relevant momenta become nearly anti-parallel, driving the response out of phase. The change from an in-phase to out-of-phase response coincides with the region where $\theta$ achieves a value of precisely $0,$ at which point the off-diagonal element $\chi_{\up\dn}$ vanishes and $\phi$ is ill-defined. We can approximate the location of this change by estimating where the non-interacting Lindhard function $\chi^0_{\up\dn}$ vanishes, which yields
\begin{align}
\label{eq:phase_switch}
        \omega = \frac{q^2}{2r_s^2}\left( -1 + \sqrt{9 - \frac{8 k_\mathrm{F}^2}{q^2}} \right)
        ,
\end{align}
in good agreement with the crossover observed in the figure, with a derivation given in Appendix~\ref{sec:phaseap}.

The response at $\lambda=1.0$, plotted in Fig.~\ref{fig:OR_response} (a.3-c.3), is quite similar to that observed at small $\lambda.$ We see that $\theta$ is pinned close to $\pi$, reflecting the fact that the skyrmion texture primarily points along the $\dn$ direction at occupied momenta for large $\lambda.$ 
The dominant eigenvalue does not match the density response as closely as the $\lambda=0.2$ case, since here there is still a small pseudospin $\up$ skyrmion core, despite the primarily $\dn$ polarization, but the physics is qualitatively unchanged compared to small $\lambda.$  

In Fig.~\ref{fig:OR_response} (a.2-c.2) we plot $\chi_+,$ $\theta,$ and $\phi$ for $r_s=1$ and $\lambda=0.5$. The skyrmion texture of the band is comparable in size to the Fermi surface for this intermediate value of $\lambda,$ resulting in starkly different results from the $\lambda=0.2$ case. First, we see the dominant eigenvalue $\chi_{+}$ now exhibits significant suppression along the centre of the particle-hole continuum, differing significantly from the total density-density response $\chi_{\rho\rho}$ that is instead suppressed along the right edge of the particle-hole continuum. Across most of the particle-hole continuum, $\theta$ takes values close to $\pi/2$, indicating that both orbitals contribute equally to the response. Only along the centre of the continuum does $\theta$ deviate from $\pi/2$, approaching $\theta=0$ for small $\omega$ and $\theta=\pi$ for large $\omega,$ both coinciding with the suppression of $\chi_+.$ Considering the relative phase between the orbitals, we see again that the response changes abruptly from in-phase to out-of-phase across the centre of the continuum. As the response is now equally distributed across both orbitals, the relative phase has a significant impact on the nature of the response: for small $q$, the in-phase response contributes an overall density response, while at large $q,$ density oscillates between the two orbitals without contributing any net density response, explaining the suppression of the density response exhibited in Fig.~\ref{fig:response}.

\section{Crystalline response functions}
In this section, we  consider the density and current response of the system in the low density limit of the phase diagram, which exhibits three distinct crystalline ground states. The presence of quantum geometry has an impact on the collective modes of these crystalline phases, enabling novel excitations not present in the WC phase of the conventional two-dimensional electron gas. We study the system in detail at a representative point for each crystalline phase: $(r_s=20,\lambda=0.2)$ for the WC, $(r_s=10,\lambda=1.6)$ for the AHC, and $(r_s=20,\lambda=2)$ for the HWC.

Before discussing the collective modes, we review how the crystalline phases of $\lambda-$jellium are distinguished at the mean-field level. These crystal phases lower their energy below that of the FL by spontaneously reorganizing their charge into a triangular lattice~\footnote{The square lattice AHC is lower in energy than the triangle lattice in a portion of the phase diagram, but this is not the case for the parameters we consider here~\cite{soejimaJelliumModelAnomalous2025}.}, as shown in Fig.~\ref{fig:HF_crystal}~(d). This density rearrangement generates a periodic potential that opens a gap in the spectrum, plotted in Fig.~\ref{fig:HF_crystal}~(a), producing a band insulator and lowering the ground state energy.

The key differences between these phases lie in their band topology and momentum space charge distributions. The WC we consider at $(r_s=20,\lambda=0.2)$ has broken time-reversal symmetry, owing to the finite value of $\lambda.$ This results in a finite Berry curvature in the occupied band, plotted in Fig.~\ref{fig:HF_crystal}~(b.1), but it is weak and integrates to zero, producing a trivial $C=0$ insulator. The broken time-reversal symmetry also manifests as a small orbital magnetization, producing circulating currents around the charge centres, as shown in Fig.~\ref{fig:HF_crystal}~(d.1). For larger values of $\lambda,$ the occupied band inherits sufficient Berry curvature from the parent band to produce a finite Chern number, resulting in a $C=1$ AHC. For the AHC at $(r_s=10,\lambda=1.6),$ the Berry curvature of the occupied band, shown in Fig.~\ref{fig:HF_crystal}~(b.2), is sharply peaked at the origin, reflecting the distribution of Berry curvature in the parent band. The AHC also hosts circulating equilibrium currents, see Fig.~\ref{fig:HF_crystal}~(d.2), but the circulations are located between the charge centres, rather than on the charge centres, and rotate in the opposite direction to those of the WC. This current distribution reflects the fact that a current-carrying chiral edge mode will form at boundary terminations of the AHC, whereas the boundary of a WC is gapped.

The HWC distinguishes itself from the WC and AHC in its momentum space charge distribution, plotted for each phase in Fig.~\ref{fig:HF_crystal}~(c). The charge distribution of the WC and AHC is concentrated in the first Brillouin zone and is peaked at the origin. The HWC instead fully depletes $k=0$ as a result of the energetic penalty to the exchange term caused by the Berry curvature concentrated there, shifting that charge into the first ring of reciprocal lattice vectors. Because the peak of the parent band Berry curvature at $k=0$ is unoccupied, the valence band of the resulting insulator has a vanishing Chern number, $C=0.$ The ground state again exhibits circulating currents around the charge centres, as shown in Fig.~\ref{fig:HF_crystal}~(d.3) but the circulation direction is opposite to that of the WC, reflecting the finite angular momentum of the ground state~\cite{soejimaJelliumModelAnomalous2025}.

\begin{figure}
    \centering
    \includegraphics[width=\linewidth]{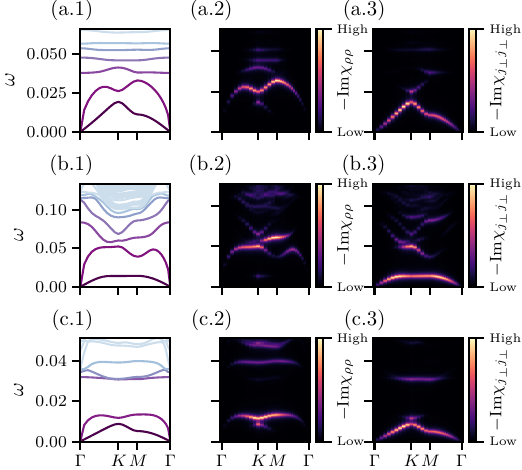}
    \caption{ The (a.1) collective mode spectrum along high-symmetry lines across the BZ, (a.2) imaginary part of the density-density response, $\chi_{\rho\rho}(\vb{q},\omega)$, and (a.3) imaginary part of the transverse current-current response, $\chi_{j_\perp j_\perp}(\vb{q},\omega),$ of the WC at $r_s=20$ and $\lambda=0.2.$ Panels (b.1-b.3) are the same for the AHC at $r_s=10$ and $\lambda=1.6,$ and panels (c.1-c.3) for the HWC at $r_s=20$ and $\lambda=2.0.$ The line colours in (a.1-b.1) depict different bands. The parameters chosen for the WC, AHC, and HWC are the same as those in Fig.~\ref{fig:HF_crystal}.
    }
    \label{fig:TDHF_crystal}
\end{figure}

Having characterized the underlying HF reference states, we now consider the collective modes, obtained via TDHF. For each parameter point considered above, we examine the collective mode spectrum, density-density response $\chi_{\rho\rho}$, and transverse current-current response $\chi_{j_\perp j_\perp}$, as plotted in Fig.~\ref{fig:TDHF_crystal}. The spectrum of each phase exhibits two gapless phonon modes, the upper and lower branches being the longitudinal and transverse phonon modes. In addition to the phonon modes, each crystal phase also possesses several gapped exciton modes that are pulled down from the particle-hole continuum by strong interactions.

Specializing to the collective mode spectrum of the WC, plotted in Fig.~\ref{fig:TDHF_crystal}~(a.1), we observe that its phonons are gapped at the $K$ point---in contrast with the $\lambda = 0$ case, in which time-reversal symmetry forces a band touching at the $K$ point~\cite{jainElementaryExcitationsMelting2025}. The density-density and transverse current-current responses, shown in Fig.~\ref{fig:TDHF_crystal}~(a.2) and (a.3), are dominated by the longitudinal and transverse phonons, respectively, with some small spectral weight carried by the excitons. The avoided crossing between the phonons at the $K$ point also leads to some of the density-density response being carried by the transverse phonon branch, and likewise for the current-current response and the longitudinal branch. The picture appears significantly more complex in the AHC, plotted in Fig.~\ref{fig:TDHF_crystal}~(b), as the longitudinal phonon now intersects the lowest exciton band. This can be seen by inspecting the density-density response, in which the spectral weight of the longitudinal phonon along the $\Gamma-K$ line continues into the lowest exciton band following the avoided crossing at the $K$ point. Besides this complication, the only other qualitative difference is that the transverse phonon of the AHC is significantly softened compared to that of the WC. The collective mode spectrum and responses of the HWC, plotted in Fig.~\ref{fig:TDHF_crystal}~(c), are also  qualitatively very similar to the WC, the only exceptions being the much larger gap between the phonons and excitons and a larger fraction of the spectral weight being carried by the excitons.

\begin{figure}
    \centering
    \includegraphics[]{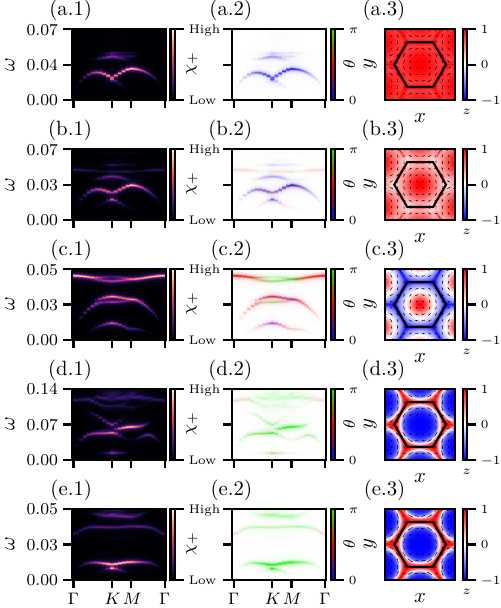}
    \caption{
    The (a.1) dominant eigenvalue, $\chi_+,$ and (a.2) orbital polarization angle, $\theta,$ of the orbital-resolved density response of the WC at $r_s = 20, \lambda=0.1$. Panel (a.3) shows the Bloch sphere projection of the real-space pseudospin texture for the same parameters, with colour indicating the $\hat{z}$ projection of the pseudospinor and arrows depicting the in-plane component. The pseudospinor is nearly completely polarized along $\hat{z},$ with no skyrmion texture. Panels (b.1-b.3) depict the same for the WC at  $r_s = 20, \lambda=0.2,$ which exhibits a weak inter-orbital mode at the $\Gamma$ point and a meron-like real-space pseudospin texture. Increasing $\lambda$ to $\lambda=0.4,$ depicted in (c.1-c.3), the ground state remains a WC but the real-space pseudospin texture now exhibits a skyrmion lattice with skyrmion charge $+1$, accompanied by a strong inter-orbital response at $\Gamma.$ The AHC at $r_s = 10, \lambda=1.6,$, shown in (d.1-d.3), also exhibits a real-space skyrmion lattice, but with skyrmion charge $-1$. The orbital-resolved density response possesses an inter-orbital mode, but it is weak compared to the WC. The HWC at $r_s = 20, \lambda=2,$ plotted in (e.1-e.3), shows the same weak inter-orbital mode and skyrmion lattice as the AHC.
    }
    \label{fig:skyrmion}
\end{figure}

Motivated by the insight gained from the orbital-resolved density-density response of the FL, we similarly study the orbital resolved density-density response of the crystalline phases Eq.~\eqref{eq:OR_matrix}. We plot in Fig.~\ref{fig:skyrmion}~(a.1-e.1) the larger of the two eigenvalues of the anti-Hermitian response tensor, $\chi_+,$ for a range of parameter points encompassing all three crystalline phases. In most cases, this eigenvalue tracks the value of the density-density response. However, for all but the smallest value of $\lambda$ considered in Fig.~\ref{fig:skyrmion}, one band of $\chi_+$ exhibits a large value near the $\Gamma$ point, where charge conservation insists that the total density-density response must vanish. In Fig.~\ref{fig:skyrmion}~(a.2-e.2), we plot the angle $\theta$ that characterizes the relative weight of the response between the two orbitals, observing that it approaches $\pi/2$ at the $\Gamma$ point wherever $\chi_+$ is finite. We further find that the relative phase, $\phi,$ between the orbitals at these points is $\pi$, indicating that this is an out-of-phase inter-orbital response, in which density shifts between the orbitals while the density per site remains fixed. 

We can understand the origin of this mode in terms of the low-energy excitations of an emergent real-space pseudospin skyrmion lattice, an idea recently put forward as a real-space explanation for the appearance of interaction-driven topological phases in RNG~\cite{tanIdealLimitRhombohedral2025,maymann2026skyrmionfractionalcherninsulator}. In short, it was shown that repulsive interactions can drive the formation of a real-space pseudospin skyrmion lattice (distinct from the momentum-space skyrmion texture of the $\lambda-$jellium model) that generates an effective magnetic field, which in turn generates a Chern band. The real-space skyrmion texture is obtained from the Fourier transform of the density matrix, the eigenvector of its largest eigenvalue serving as the real-space pseudospinor, as discussed in Appendix~(C.8) in reference~\cite{tanIdealLimitRhombohedral2025}. We discuss in the following paragraph how the appearance of the real-space pseudospin texture coincides with the appearance of an out-of-phase inter-orbital response at the $\Gamma$ point in the crystal phases of $\lambda-$jellium.

In $\lambda-$jellium, the real-space pseudospinor is two-dimensional and maps onto the Bloch sphere, allowing us to plot the real-space pseudospin texture in Fig.~\ref{fig:skyrmion}~(a.3-e.3) using arrows to represent the in-plane component and colour to depict the projection onto the $\hat{z}$ axis. For the WC at $(r_s, \lambda)=(20, 0.1),$ plotted in Fig.~\ref{fig:skyrmion} (a.3), the pseudospin texture points along the $+\hat{z}$ axis across the unit cell, reflecting the small value of $\lambda,$ and exhibits no inter-orbital response at the $\Gamma$ point in momentum space. In contrast, the WC at $(r_s, \lambda)=(20, 0.2)$ plotted in Fig.~\ref{fig:skyrmion}~(b.3) shows a meron-like texture in real space, with the pseudospinor pointing along $+\hat{z}$ at centre of the unit cell and pointing in-plane at the unit cell boundary. The orbital-resolved density response here, plotted in Fig.~\ref{fig:skyrmion}~(b.2), also possesses a weak inter-orbital mode at $\Gamma.$ For $(r_s, \lambda)=(20, 0.4),$ plotted in Fig.~\ref{fig:skyrmion} (c.3), the real-space pseudospin texture of the WC becomes a skyrmion lattice with skyrmion charge $+1$, and the inter-orbital mode at the $\Gamma$ point becomes quite strong in the density response, shown in Fig.~\ref{fig:skyrmion}~(c.2). The AHC at $(r_s, \lambda)=(10, 1.6)$ and the HWC at $(r_s, \lambda)=(20, 2),$ plotted in Figs.~\ref{fig:skyrmion}~(d) and (e), respectively, also both exhibit a real-space skyrmion lattice, but with the opposite skyrmion charge as that of the WC. The orbital-resolved density response of AHC and HWC also exhibit the inter-orbital mode at $\Gamma$, although it is not as strong as in the WC. These out-of-phase inter-orbital modes correspond to the breathing mode excitation of the skyrmion lattice, in which charge oscillates between orbitals with no change in the total density, as discussed in~\cite{tanIdealLimitRhombohedral2025}.

\section{Discussion} 
In this work, we applied the TDHF approach to examine the impact of quantum geometry on the liquid--crystal phase transition of the two-dimensional electron gas. Our calculations confirm the recently reported trend that quantum geometry favours crystallization, significantly increasing the critical density of the transition~\cite{valentiQuantumGeometryDriven2025}. Our TDHF calculations also revealed a novel instability of the FL ground state near the liquid--crystal transition, in which strong interactions and concentrated Berry curvature drive fluctuations between states deep within the Fermi sea and states just outside Fermi surface. This instability emerges nearby the liquid--crystal transition, indicating that associated fluctuations likely encourage crystallization. We hypothesize that these fluctuations may be the reason why the liquid state struggles to generate an energy-efficient exchange-correlation hole, as discussed in~\cite{valentiQuantumGeometryDriven2025}, tipping the energetic competition in favour of the crystal. We leave further investigation of this hypothesis to future work.

Our investigation of the current and density response of the Fermi liquid demonstrated that its collective excitations are also significantly affected by quantum geometry. Large regions of the density response are suppressed for intermediate values of $\lambda$ and Friedel oscillations are completely suppressed at $\lambda = 0.5$, for which the relevant scattering states on the Fermi surface are orthogonal. This phenomenon is similar to the quantum geometric Friedel oscillations reported in a recent paper, although the microscopic origin of the effect is quite different~\cite{ma2026quantumgeometricfriedeloscillations}.

Through a combination of numerics and analytic arguments we also showed that the dispersion of the plasmon mode is reduced by the presence of a finite quantum metric. Resolving the density-density response orbital by orbital, we showed that the suppressed density response is replaced by an out-of-phase inter-orbital mode in which the total density remains fixed, but density oscillates between orbitals. Applying a similar analysis to the crystal phases at low densities, we observed a similar out-of-phase inter-orbital mode appear at the $\Gamma$ point across wide swaths of the phase diagram. Computing the real-space pseudospin structure, we identified this mode with the breathing mode of an emergent real-space skyrmion lattice~\cite{tanIdealLimitRhombohedral2025, maymann2026skyrmionfractionalcherninsulator}.

\begin{acknowledgments}
We thank F\'elix Desrochers for illuminating conversations. This work is supported by the Natural Sciences and Engineering Research Council of Canada (NSERC) and the Centre for Quantum Materials at the University of Toronto. P.F. was further supported by the NSERC Canada Graduate Scholarships-Doctoral (CGS-D). Computations were performed on the Fir cluster, hosted by the Digital Research Alliance of Canada.
\end{acknowledgments}

\appendix

\section{Instability towards annular Fermi liquid ground state}
\label{sec:AFLap}

The TDHF stability analysis performed in Section \ref{sec:pd} of the main text reveals a novel instability enabled by quantum geometry, corresponding to particle-hole excitations that eject electrons from the Brillouin zone centre to the Fermi surface. This tendency is similar to the physics that underlies the annular Fermi liquid, in which the occupied states form an annulus to avoid the significant energetic penalty imposed by the strong Berry curvature at small momenta. However, the conventional FL is lower in energy than the annular FL across the entire phase diagram we consider in the main text, so the instability we observe must be towards some other competing ground state.

\begin{figure}[b]
    \centering
    \includegraphics[width=1.0\linewidth]{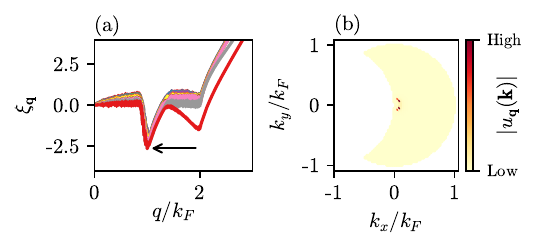}
    \caption{(a) The lowest ten eigenvalues of the stability matrix $S_{\vb{q}}$, for the point $(r_s,\lambda) = (10,3)$, with different bands shown in distinct colours. At this point in the phase diagram, the annular FL is known to be lower in energy than the reference regular FL state. This is reflected by the dominant instability at $q=k_\mathrm{F}$, present for all bands plotted. (b) The eigenvector of the lowest band at $q=k_F$, as depicted by the arrow.}
    \label{fig:AFL}
\end{figure}

Given the similarities between the observed instability and the physics of the annular FL, it is natural to also perform a TDHF stability analysis on the FL phase in a region where the AFL is lower in energy. In Fig.~\ref{fig:AFL} we present such an analysis for the FL HF ground state at the point $(r_s = 10, \lambda = 3),$ where the AFL is lower in energy~\cite{soejimaJelliumModelAnomalous2025}. In Fig.~\ref{fig:AFL}~(a), we plot the 10 lowest eigenvalues of the stability matrix as a function of $\vb{q}/k_{\mathrm{F}}$. We see that many bands become negative around $q=k_{\mathrm{F}}$, and, despite the presence of an instability at $q=2k_{\mathrm{F}},$ the most negative values appear at $q=k_{\mathrm{F}}.$ The eigenvector of the lowest band at $q=k_\mathrm{F}$, plotted in Fig.~\ref{fig:AFL}~(b), is sharply peaked at a few isolated momentum points near the origin, with similar structure found for the eigenvectors of the other negative bands at $q=k_\mathrm{F}$. These features distinguish the FL--AFL instability from those discussed in the main text, each of which are qualitatively distinct.

\section{Plasmon dispersion within RPA}
\label{sec:RPAap}
In this section, we derive the quantum metric correction to the energy of the plasmon mode within the random phase approximation (RPA). Within the RPA~\cite{Giuliani_Vignale_2005}, the density-density response function is given by 
\begin{align}
    \chi^\mathrm{RPA}_{\rho \rho}(\vb{q},\omega) = \frac{\chi^{0}_{\rho\rho}(\vb{q},\omega)}{1 - V(q)\chi^{0}_{\rho\rho}(\vb{q},\omega)}
    ,
\end{align}
where here $\chi^0_{\rho\rho}$ is the non-interacting Lindhard response function,
\begin{align}
\chi_{\rho\rho}^0(\vb{q},\omega) = \frac{1}{A}\sum_{\vb{k}} |\mathcal{F}(\vb{k},\vb{k}+\vb{q})|^2 \frac{n_{\vb{k}} - n_{\vb{k} + \vb{q}}}{\omega + \varepsilon_{\vb{k}} - \varepsilon_{\vb{k}+\vb{q}}}
.
\end{align}
The dispersion of the plasmon mode is given by the pole of $\chi_{\rho\rho}^\mathrm{RPA}(\vb{q},\omega)$, namely, by the relation
\begin{align}
    \label{eq:pole}
    1 = V(q) \chi_{\rho\rho}^0(\vb{q},\omega)
    .
\end{align}
We solve for the dispersion in the small $q$ limit and assume isotropy throughout, expanding the form factor as $|\mathcal{F}(\vb{k},\vb{k}+\vb{q})|^2 \approx 1 - \frac{q^2 \tr g(\vb{k})}{2}$. Here $g$ is the quantum metric, with matrix elements $g_{ab}(\vb{k}) = \operatorname{Re} \mel{ \partial_{k_a} u_{\vb{k}}}{\left( \mathds{1} - \ket{u_{\vb{k}}} \bra{u_{\vb{k}}} \right)}{\partial_{k_b} u_{\vb{k}}}$. To expand the Lindhard function, we first rewrite the sum in the following way:
\begin{widetext}
\begin{align}
        \chi_{\rho\rho}^0(\vb{q},\omega) &= \frac{1}{A}\sum_{\vb{k}} |\mathcal{F}(\vb{k},\vb{k}+\vb{q})|^2 \frac{n_{\vb{k}}}{\omega + \varepsilon_{\vb{k}} - \varepsilon_{\vb{k}+\vb{q}}} - \frac{1}{A}\sum_{\vb{k}} |\mathcal{F}(\vb{k},\vb{k}+\vb{q})|^2 \frac{n_{\vb{k} + \vb{q}}}{\omega + \varepsilon_{\vb{k}} - \varepsilon_{\vb{k}+\vb{q}}} \\
        &= \frac{1}{A}\sum_{\vb{k}} |\mathcal{F}(\vb{k},\vb{k}+\vb{q})|^2 \frac{n_{\vb{k}}}{\omega + \varepsilon_{\vb{k}} - \varepsilon_{\vb{k}+\vb{q}}}
        - 
        \frac{1}{A}\sum_{\vb{k}} |\mathcal{F}(-\vb{k}-\vb{q},-\vb{k})|^2 \frac{n_{-\vb{k}}}{\omega + \varepsilon_{-\vb{k}-\vb{q}} - \varepsilon_{-\vb{k}}}  \\
        &= \frac{1}{A}\sum_{\vb{k}} |\mathcal{F}(\vb{k},\vb{k}+\vb{q})|^2 n_{\vb{k}} \left(\frac{1}{\omega + \varepsilon_{\vb{k}} - \varepsilon_{\vb{k}+\vb{q}}} + (\omega \longleftrightarrow -\omega) \right)
        \label{eq:relabel}
    \end{align}
We next expand to sixth order in $q,$ obtaining
\begin{align}
    \chi_{\rho\rho}^0(\vb{q},\omega) \approx \frac{2}{A}\sum_{\vb{k}}n_{\vb{k}} 
    \left( \frac{q^2}{r_s^2 \omega^2} + \frac{12 (\vb{k}\cdot\vb{q})^2q^2}{r_s^6 \omega^4} + \frac{80 (\vb{k}\cdot\vb{q})^4 q^2}{r_s^{10}\omega^6} - \frac{q^4 \tr g(\vb{k})}{2r_s^2 \omega^2} \right) 
\end{align}
Changing the sum to an integral and evaluating each term, we arrive at
\begin{align}
    \chi_{\rho\rho}^0 \approx \frac{q^2 k_\mathrm{F}^2}{2\pi r_s^2 \omega^2}+\frac{3q^4k_\mathrm{F}^4}{2\pi r_s^6 \omega^4} + \frac{5 q^6 k_\mathrm{F}^6}{\pi r_s^{10} \omega^6} - \frac{q^4}{4 \pi^2 r_s^2 \omega^2}\int_{k<k_\mathrm{F}} d^2 \vb{k} \tr g(\vb{k}).
\end{align}
Substituting this expression into Eq.~\eqref{eq:pole} and inserting the Coulomb interaction, we find
\begin{align}
    \omega^2 \approx \frac{2q k_\mathrm{F}^2}{r_s^3} + \frac{6 q^3 k_\mathrm{F}^4}{r_s^7\omega^2} + \frac{20 q^5 k_\mathrm{F}^6}{r_s^{11}\omega^4} - \frac{q^3}{\pi r_s^3} \int_{k<k_\mathrm{F}} d^2 \vb{k} \tr g(\vb{k}).
    \label{eq:rpa_exp}
\end{align}
In the limit of small $q$, we can substitute the expansion $\omega^2 \approx Aq + Bq^2 + Cq^3$ on both sides of Eq.~\eqref{eq:rpa_exp} and, matching powers, obtain 
\begin{align}
    \omega^2 \approx \frac{2qk_\mathrm{F}^2}{r_s^3} + \frac{3q^2k_\mathrm{F}^2}{r_s^4} + \frac{q^3k_\mathrm{F
    }^2}{2r_s^5} - \frac{q^3}{\pi r_s^3} \int_{k<k_\mathrm{F}} d^2 \vb{k} \tr g(\vb{k}).
\end{align}
Recalling $k_\mathrm{F}=2$ and introducing the notation $\omega_\mathrm{p}^0 = \frac{2}{r_s} \sqrt{\frac{2 q}{r_s}}$ for the classical plasmon dispersion of the conventional two-dimensional electron gas, we finally obtain
\begin{align}
    \omega \approx \omega_\mathrm{p}^0\left(1 + \frac{3q}{4r_s} - \frac{5q^2}{32r_s^2} - \frac{q^2}{4\pi k_\mathrm{F}^2} \int_{k<k_\mathrm{F}} d^2 \vb{k} \tr g(\vb{k})\right).
\end{align}
Expressed in terms of density-independent units, this takes the form
\begin{align}
    \omega \approx \omega_\mathrm{p}^0\left(1 + \frac{3qa_B}{4} - \frac{5q^2a_B^2}{32} - \frac{q^2}{4\pi k_\mathrm{F}^2} \int_{k<k_\mathrm{F}} d^2 \vb{k} \tr g(\vb{k})\right),
\end{align}
with $\omega_\mathrm{p}^0 = \frac{2 \sqrt{2 q a_B}}{r_s}~\mathrm{Ry}$. The geometric contribution discussed in the main text is given by the difference between the dispersion with for finite $\lambda$, $\omega^\mathrm{RPA,\lambda}_{\mathrm{p}}$, and for $\lambda = 0$, $\omega^\mathrm{RPA,0}_{\mathrm{p}}$:
\begin{align}
\omega^\mathrm{RPA,\lambda}_{\mathrm{p}} - \omega^\mathrm{RPA,0}_{\mathrm{p}}   \sim -\frac{\omega^0_\mathrm{p}q^2}{4\pi k_\mathrm{F}^2}\int_{k < k_\mathrm{F}}d^2 \vb{k} \tr g(\vb{k})
\end{align}

\section{In-phase to out-of-phase transition in orbital-resolved density response}
\label{sec:phaseap}
Here we compute the boundary between the in-phase to out-of-phase orbital-resolved density-density response of the FL, as seen in Fig.~\ref{fig:OR_response} and described by Eq. (\ref{eq:phase_switch}) in the main text. The phase of the orbital-resolved response is determined by the sign of the off-diagonal component of the anti-Hermitian response tensor, $\hat{\chi}_{\mathrm{AH}}$. Physically, this component describes density response of the pseudospin $\up$ orbital to a perturbation on the $\dn$ orbital. The response is in-phase when $[\hat{\chi}_{\mathrm{AH}}]_{\up \dn} >0$ and out-of-phase when $[\hat{\chi}_{\mathrm{AH}}]_{\up \dn} <0$, switching where the $[\hat{\chi}_{\mathrm{AH}}]_{\up \dn}$ vanishes.

To obtain an analytical approximation, we take the non-interacting limit and compute where 
\begin{align}
    [\hat{\chi}_{\mathrm{AH}}]_{\up\dn}
    &\approx
    \frac{1}{2i}\left(\chi^0_{\up\dn} - \chi^{0*}_{\dn\up}\right)
    \\
    &=
    \frac{1}{A}\sum_{\vb{k}}
    \mathcal{F}_{\up}(\vb{k},\vb{k}+\vb{q})\mathcal{F}_{\dn}(\vb{k}+\vb{q},\vb{k}) (n_{\vb{k}} - n_{\vb{k} +\vb{q}})\left[\frac{1}{\omega + \varepsilon_{\vb{k}} - \varepsilon_{\vb{k} + \vb{q}} + i0^+} - \frac{1}{\omega + \varepsilon_{\vb{k}} - \varepsilon_{\vb{k} + \vb{q}} - i0^+}\right]
\end{align}
vanishes, utilizing the following expression for the non-interacting inter-orbital Lindhard function:
\begin{align}
   \chi_{\up\dn}^0(\vb{q},\omega) = \frac{1}{A}\sum_{\vb{k}}\mathcal{F}_{\up}(\vb{k},\vb{k}+\vb{q})\mathcal{F}_{\dn}(\vb{k}+\vb{q},\vb{k}) \frac{n_{\vb{k}} - n_{\vb{k}+\vb{q}}}{\omega + \varepsilon_{\vb{k}} - \varepsilon_{\vb{k}+\vb{q}} + i0^+}
   .
\end{align}
Applying the Sokhotski–Plemelj theorem, $\frac{1}{x+i0^+} = \mathcal{P}(\frac{1}{x}) - i\pi\delta(x)$, the expression simplifies to
\begin{align}
     \frac{1}{2i}\left(\chi^0_{\up\dn} - \chi^{0*}_{\dn\up}\right) = 
    -\frac{\pi}{A}\sum_{\vb{k}} \mathcal{F}_{\up}(\vb{k},\vb{k}+\vb{q})\mathcal{F}_{\dn}(\vb{k}+\vb{q},\vb{k})(n_{\vb{k}} - n_{\vb{k} + \vb{q}})\delta(\omega + \varepsilon_{\vb{k}} - \varepsilon_{\vb{k}+\vb{q}})
    .
\end{align}
Performing an analogous relabelling to the one done in Eq.~\eqref{eq:relabel}, and changing the sum to an integral, we arrive at 
\begin{align}
    \frac{1}{2i}\left(\chi^0_{\up\dn} - \chi^{0*}_{\dn\up}\right)
    &=
    -\frac{1}{4\pi} \int_{|\vb{k}|\leq k_\mathrm{F}} d^2\vb{k}  \mathcal{F}_{\up}(\vb{k},\vb{k}+\vb{q})\mathcal{F}_{\dn}(\vb{k}+\vb{q},\vb{k}) \delta(\omega + \varepsilon_{\vb{k}} - \varepsilon_{\vb{k}+\vb{q}})
    \\
    &+
    \frac{1}{4\pi} \int_{|\vb{k}|\leq k_\mathrm{F}} d^2\vb{k}  \mathcal{F}_{\up}(\vb{k}+\vb{q}, \vb{k})\mathcal{F}_{\dn}(\vb{k}, \vb{k}+\vb{q}) \delta(\omega - \varepsilon_{\vb{k}} + \varepsilon_{\vb{k}+\vb{q}})
    .
\end{align}
For $\omega,q>0$, the $\delta$-function in the second term vanishes wherever $\omega> q/r_s(2k_\mathrm{F}-q)$. We restrict our estimate to this region, which constitutes the majority of the particle-hole continuum. We therefore proceed by determining where the first term vanishes. Explicitly evaluating the form factors and changing to polar coordinates, the first term can be written as
\begin{align}
    -\frac{\lambda^2}{4\pi} \int_0^{k_\mathrm{F}} dk \int_0^{2\pi} d\theta \frac{k(k^2 + kq\cos\theta - ikq\sin\theta)}{(1+ \lambda^2k^2)(1+\lambda^2(k^2 + q^2 + 2kq\cos \theta))} \delta(\omega - q^2/r_s^2 - 2kq\cos\theta/r_s^2).
\end{align}
Leveraging the $\delta$-function to replace each instance of $kq\cos\theta,$ we find
\begin{align}
    -\frac{\lambda^2}{4\pi} \int_0^{k_\mathrm{F}} dk\int_0^{2\pi} d\theta  \frac{k(k^2 + \omega r_s^2/2 - q^2/2 - ikq\sin\theta)}{(1+ \lambda^2k^2)(1+\lambda^2(k^2 + \omega r_s^2))}  \delta(\omega - q^2/r_s^2 - 2kq\cos\theta/r_s^2)
\end{align}
We note that the $\sin \theta$ term vanishes by symmetry, since $\int_0^{2\pi}  \sin\theta F(\cos(\theta)) d\theta = \int_{-\pi}^{\pi}  \sin\theta F(\cos(\theta)) d\theta$ is clearly odd for any function $F$. We are left to evaluate
\begin{align}
    -\frac{\lambda^2}{4\pi} \int_0^{k_\mathrm{F}} dk  \frac{k(k^2 + \omega r_s^2/2 - q^2/2 )}{(1+ \lambda^2k^2)(1+\lambda^2(k^2 + \omega r_s^2))} \int_0^{2\pi} d\theta \delta(\omega - q^2/r_s^2 - 2kq\cos\theta/r_s^2).
\end{align}
We first perform the angular integral. In order for the $\delta$-function to be non-vanishing, we require 
\begin{align}
    \left|\frac{\omega r_s^2 - q^2}{2q} \right| \leq k
    .
\end{align}
If the condition is satisfied, there are two solutions where the $\delta$-function is satisfied, one at $\theta_1 = \arccos\left(\frac{\omega r_s^2 -q^2}{2kq}\right)$ and one at $\theta_2 = 2\pi - \theta_1$. Using the property $\delta(f(\theta)) = \sum_i \frac{\delta(\theta - \theta_i)}{|f'(\theta_i)|}$, the angular integral evaluates to
\begin{align}
    \int_0^{2\pi} d\theta \delta(\omega - q^2/r_s^2 - 2kq\cos\theta/r_s^2) 
    =
    \frac{r_s^2}{2kq|\sin \theta_1|} + \frac{r_s^2}{2kq|\sin \theta_2|}
    .
\end{align}
Since $|\sin \theta_1 |= |\sin \theta_2| = \sqrt{1 - \frac{(\omega r_s^2 - q^2)^2}{4k^2q^2}}$, the angular integral becomes
\begin{align}
    \frac{2 r_s^2}{\sqrt{4k^2q^2 - (\omega r_s^2 - q^2)^2}}
    ,
\end{align}
with $k \geq \left|\frac{\omega r_s^2 - q^2}{2q} \right|$. This leaves us with the radial integral 
\begin{align}
    - \frac{\lambda^2 r_s^2}{2\pi}\int_{|\omega r_s^2 -q^2|/2q}^{k_\mathrm{F}}
    dk 
    \frac{k(k^2 + \omega r_s^2/2 - q^2/2 )}{(1+ \lambda^2k^2)(1+\lambda^2(k^2 + \omega r_s^2))\sqrt{4k^2q^2 - (\omega r_s^2 - q^2)^2}}.
\end{align}
To arrive at a simple analytical estimate for the boundary, we assume $\lambda$ is small and keep only the lowest order $(\lambda^2)$ term, yielding 
\begin{align}
    \approx
    - \frac{\lambda^2 r_s}{2\pi}\int_{|\omega r_s^2 -q^2|/2q}^{k_\mathrm{F}}
    dk 
    \frac{k(k^2 + \omega r_s^2/2 - q^2/2 )}{\sqrt{4k^2q^2 - (\omega r_s^2 - q^2)^2}}
\end{align}
To evaluate the remaining integral, we define $A = (\omega r_s^2 -q^2)/2q$ and recast it as
\begin{align}
    - \frac{\lambda^2 r_s^2}{2\pi}\int_{|A|}^{k_\mathrm{F}}
    dk 
    \frac{k(k^2 +qA)}{2q\sqrt{k^2 - A^2}}
    .
\end{align}
Making the substitution $u = k^2 - A^2$, we get
\begin{align}
    -\frac{\lambda^2 r_s^2}{8\pi q} \int_0^{k_\mathrm{F}^2 - A^2} du (u^{1/2} + u^{-1/2}( A^2 + qA))
\end{align}
Evaluating the integral, we arrive at the result
\begin{align}
    -\frac{\lambda^2 r_s^2}{4 \pi q}\sqrt{k_\mathrm{F}^2 - A^2}\left( \frac{1}{3}k_\mathrm{F}^2 + \frac{2}{3}A^2 + q A \right),
\end{align}
and the phase boundary coincides with where the term in parentheses vanishes. Substituting $A = (\omega r_s^2 -q^2)/2q$ back in and solving the quadratic equation, we arrive at 
\begin{align}
    \omega = \frac{q^2}{2r_s^2}\left( -1 \pm \sqrt{9 - \frac{8k_\mathrm{F}^2}{q^2}} \right)
\end{align}
The relevant solution for the observed boundary from in phase to out of phase response is the $+$ branch, recovering Eq. (\ref{eq:phase_switch}) given in the main text. This solution is valid when $q/k_\mathrm{F} \geq 2 \sqrt{2}/3$ and $\omega> q/r_s(2k_\mathrm{F}-q)$ both hold. Deviations between this estimate and the boundary observed in Fig.~\ref{fig:OR_response} of the main text arise from the fact that we have ignored interactions and expanded to lowest order in $\lambda$. 
\end{widetext}
\bibliography{main.bib}
\end{document}